\documentclass[pra,ams,aps,twocolumn,superscriptaddress,nofootinbib]{revtex4}

\input Qcircuit
\input epsf

\usepackage{graphicx}

\usepackage{amsmath} \usepackage{amsfonts} \usepackage{latexsym}
\usepackage{ifthen}

\def\ad{\mbox{ad}}
\def\tr{\mbox{tr}}
\def\wt{\mbox{wt}}

\newtheorem{theorem}{Theorem}
\newtheorem{corollary}{Corollary}
\newtheorem{proposition}{Proposition}

\begin{document}

\title{The geometry of quantum computation}
\author{Mark R. Dowling}
\affiliation{School of Physical Sciences, The University of
Queensland, Brisbane, Queensland 4072, Australia}
\author{Michael A. Nielsen} \thanks{nielsen@physics.uq.edu.au and
  www.qinfo.org/people/nielsen}
\affiliation{School of Physical Sciences, The University of
Queensland, Brisbane, Queensland 4072, Australia}
\date{\today}

\begin{abstract}
  Determining the quantum circuit complexity of a unitary operation is
  closely related to the problem of finding minimal length paths in a
  particular curved geometry [Nielsen \emph{et al}, Science
  \textbf{311}, 1133-1135 (2006)].  This paper investigates many of
  the basic geometric objects associated to this space, including the
  Levi-Civita connection, the geodesic equation, the curvature, and
  the Jacobi equation.  We show that the optimal Hamiltonian evolution
  for synthesis of a desired unitary necessarily obeys a simple
  universal geodesic equation.  As a consequence, once the initial
  value of the Hamiltonian is set, subsequent changes to the
  Hamiltonian are completely determined by the geodesic equation.  We
  develop many analytic solutions to the geodesic equation, and a set
  of invariants that completely determine the geodesics.  We
  investigate the problem of finding minimal geodesics through a
  desired unitary, $U$, and develop a procedure which allows us to
  deform the (known) geodesics of a simple and well understood metric
  to the geodesics of the metric of interest in quantum computation.
  This deformation procedure is illustrated using some three-qubit
  numerical examples.  We study the computational complexity of
  evaluating distances on Riemmanian manifolds, and show that no
  efficient classical algorithm for this problem exists, subject to
  the assumption that good pseudorandom generators exist.  Finally, we
  develop a canonical extension procedure for unitary operations which
  allows ancilla qubits to be incorporated into the geometric approach
  to quantum computing.
\end{abstract}

\maketitle

\section{Introduction}

A central problem of quantum computation is to find efficient quantum
circuits to synthesize desired unitary operations.  These unitary
operations are used to solve computational problems such as
factoring~\cite{Shor94a,Shor97a,Nielsen00a}.  Despite intensive
effort, few general principles are known either for finding efficient
quantum circuits or for proving that a given computational problem has
no efficient circuit.

A geometric approach to quantum circuit complexity has recently been
developed in~\cite{Nielsen06c,Nielsen06b,Nielsen06d}.  The idea is to
introduce a Riemannian metric on the space of $n$-qubit unitary
operations, chosen in such a way that the metric distance $d(I,U)$
between the identity operation and a desired unitary $U$ is equivalent
(modulo some technical caveats, discussed below) to the number of
quantum gates required to synthesize $U$.  Thus the distance $d(I,U)$
is a good measure of the difficulty of synthesizing $U$.

This geometric reformulation suggests that the tools of Riemannian
geometry may be useful in analyzing quantum circuit complexity.  The
purpose of this paper is to develop in detail many of the basic
geometric notions that can be associated to quantum computation,
including the Levi-Civita connection, geodesics, geodesic invariants,
curvature, Jacobi fields and conjugate points.  We also discuss
obstacles to the use of geometric ideas to analyze quantum circuit
complexity.

\textbf{Structure and content of the paper:}
Section~\ref{sec:background} reviews the relevant background results
connecting geometry and quantum circuit complexity,
from~\cite{Nielsen06c,Nielsen06d,Nielsen06b}.

Section~\ref{sec:connection} derives simple formulae for, and several
basic properties of, the Levi-Civita connection on our Riemannian
manifold.  With these results in hand all other geometric quantities
computed later in the paper may be obtained through relatively
straightforward computations.  In particular,
Appendix~\ref{sec:curvature} computes all the natural curvature
quantities for the manifold, including the curvature tensor, the
sectional curvature, the Ricci tensor, and the scalar curvature.

Section~\ref{sec:geodesics} uses the connection to derive the geodesic
equation.  This is a simple and (we believe) rather elegant equation
that determines the locally optimal Hamiltonian evolution for
synthesis of any desired unitary. We obtain a complete set of
constants of the motion for the geodesic equation, as well as many
simple exact solutions, including a completely general exact solution
for three qubits.

Section~\ref{sec:geodesic_deformation} develops a procedure for
numerically finding geodesics passing through a desired endpoint,
i.e., a goal unitary.  We begin with a review of the theory of Jacobi
fields and conjugate points, which make use of the curvature to study
the divergence or convergence of geodesics on a manifold.  These tools
can be used to study when a geodesic is no longer globally minimizing,
but is merely a local minimum, and we briefly digress to investigate
this phenomenon numerically for a class of unitaries associated with
the transverse Ising model.  Returning to the main point of the
section, finding geodesics to a goal unitary, the basic idea is to
smoothly deform the geodesics of a simple and well understood metric
to the geodesics of the metric of computational interest.  This is
done using a notion of a geodesic derivative along a flow through the
space of metrics. The idea is to fix the geodesic endpoint, and then
the geodesic derivative describes how the initial velocity of the
geodesic must change as the metric is changed, in order that the
deformed geodesic passes through the same endpoint.  We develop
necessary and sufficient conditions for the geodesic derivative to
exist, and a formula for it when it does exist.  These necessary and
sufficient conditions are naturally expressed in terms of the
conjugate points studied earlier in this section.  We then use this
deformation procedure to numerically find geodesics passing through
some unitary operations of interest, including randomly chosen
unitaries, and the quantum Fourier transform.

In Section~\ref{sec:discussion} we discuss the general prospects for
using geometric ideas to analyze quantum circuit complexity.  We
describe two important obstacles to using geometric ideas to prove
nontrivial upper or lower bounds on quantum circuit complexity, and
prove two related technical results.  The first obstacle is the
Razborov-Rudich theorem, a well-known result in classical
computational complexity.  This is essentially a no-go theorem that,
subject to certain assumptions, rules out a wide class of approaches
to proving circuit lower bounds.  We outline a quantum analogue of the
Razborov-Rudich theorem, and use it to argue that the \emph{general}
problem of finding geodesics on a Riemannian manifold is likely to
have no efficient (classical) solution. This suggests that proving
complexity lower bounds using geometric techniques will require us to
use non-generic properties of specific unitaries. The second obstacle
discussed in this section is an apparent shortcoming in prior work on
the geometric approach to quantum computing, which is that it was
developed for the analysis of quantum circuits which do not make use
of ancillary working qubits.  We show how to avoid this restriction by
using a canonical extension procedure for unitary operations which
allows ancillas to be incorporated into the geometric point of view.
This canonical extension procedure may be of independent interest.

\textbf{Background:} We assume throughout that the reader is familiar
with quantum circuits at the level of, e.g., Chapter~4
of~\cite{Nielsen00a}, and with elementary Riemannian geometry, at the
level of, e.g.,~\cite{Lee97a,doCarmo92a}.  In particular, we assume a
working knowledge of notions such as tensor fields, the Levi-Civita
connection, the geodesic equation, and the curvature tensor.  Much of
our presentation is concerned with properties of a special type of
Riemannian manifold, known as a \emph{right-invariant} manifold.
However, the results we need about right-invariant Riemannian
manifolds are not easily accessible in a single (or even a few)
publications, so far as we are aware.  Therefore, to make the paper
accessible, we have developed in a self-contained way the main results
about right-invariant manifolds.  The reader curious to investigate
the literature further should be warned that both left- and
right-invariant manifolds are widely studied, but differ only
trivially, and results about one can always be transformed into
results about the other, sometimes with changes of sign.

\textbf{Prior work:} Geometric techniques have been used previously in
the study of quantum information processing.  In particular, Khaneja
\emph{et al}~\cite{Khaneja01a,Khaneja01b,Khaneja02a} (see also related
ideas in~\cite{Khaneja06a}) have used powerful techniques from the
theory of symmetric spaces to study time-optimal control.  This has
been extremely successful in the two-qubit case, leading to an
essentially complete characterization of two-qubit time optimal
control.  In the many-qubit scenario some successes have been
achieved, but the need for the rather specialized symmetric space
structure limits the breadth of possible applications.  Related ideas
have also been investigated by Carlini \emph{et
  al}~\cite{Carlini06a,Carlini06b}, who developed variational
principles for time-optimal synthesis of quantum states and of unitary
transformations.  Time-optimal quantum control of unitary operations
has a long history; in addition to the above references, we refer the
reader to~\cite{Boscain05a,SchulteHerbruggen05a} for recent work, and
references to earlier literature.

\textbf{Perspective:} The geometric approach to quantum computation is
in its infancy, and its long-term merits remain to be determined.
This and earlier papers do not yet offer a killer application of
geometric ideas, not available through conventional circuit analysis.
However, we believe that there are reasons to hope that the geometric
viewpoint will eventually enable insights difficult to obtain in the
conventional circuit picture.  In particular, by recasting the problem
of quantum circuit complexity in terms of smooth mathematical objects,
we bring the principles of the calculus of variations into play.  This
allows us to derive a geodesic equation describing the locally optimal
evolutions generating a desired unitary; intuitively, this is to fall
freely along a minimal geodesic, with the motion determined entirely
by the local geometry of the space, and the initial direction of
motion.  This is in contrast to the circuit picture, where no such
principles are available to derive the locally optimal way to
construct a circuit.  The present paper develops a fairly complete
picture of the basic geometry of quantum computation, and lays the
foundation for a more detailed understanding.

\section{Quantum circuit complexity and geometry}
\label{sec:background}

In this section we review the connections between Riemannian geometry
and quantum circuit complexity, as developed
in~\cite{Nielsen06c,Nielsen06d,Nielsen06b}.

We begin by recalling a few basic ideas from Riemannian geometry.  We
will actually consider Riemannian metrics on two slightly different
manifolds, the group $M = U(2^n)$ of $n$-qubit unitary operations, and
the group $M = SU(2^n)$ of $n$-qubit unitaries with unit determinant.
For the most part the development is identical for the two cases, and
we will not explicitly delineate them.  However, there are a few cases
where it is advantageous to use one or the other, and we mention
explicitly when this is the case. For definiteness, you may assume
that we are working with $SU(2^n)$ unless otherwise specified.

A tangent vector to a point on this manifold (i.e., an $n$-qubit
unitary, $U$, with unit determinant) can be thought of as a traceless
Hamiltonian, $H$, i.e., an element of the Lie algebra $su(2^n)$ of
traceless $2^n \times 2^n$ Hermitian matrices.  More precisely, we can
identify a tangent vector at $U$ with the tangent to the curve
$e^{-iHt}U$ at $t=0$.  We shall call $H$ the \emph{Hamiltonian
  representation} of this tangent vector\footnote{In an earlier
  paper~\cite{Nielsen06b} the Pauli expansion coefficients of $H$ were
  referred to as local co-ordinates for the tangent vector.  We shall
  not use this terminology here.}.  With these identifications, the
Riemannian metric $\langle \cdot, \cdot \rangle_U$ at a point $U$ is a
positive-definite bilinear form $\langle H, J \rangle_U$ defined on
traceless Hamiltonians $H$ and $J$.  Through most of this paper we
assume that this bilinear form is constant as a function of $U$ and so
write $\langle \cdot, \cdot \rangle_U = \langle \cdot,\cdot \rangle$.
A metric which is constant in this way is known as a
\emph{right-invariant} metric.

Suppose $U(t)$ is a curve in $SU(2^n)$ generated by the Hamiltonian
$H(t)$ according to the Schr\"odinger equation $\dot U = -i H U$.
Then the length of that curve is given by
\begin{eqnarray}
  \int dt \, \langle H(t), H(t)\rangle^{1/2}.
\end{eqnarray}
The distance $d(U,V)$ between points $U$ and $V$ in $SU(2^n)$ is
defined to be the minimal length of any curve joining those two
points.

The metric of most interest in this paper is defined as follows.  Let
${\cal P}$ be the subspace of $n$-qubit Hamiltonians which contain
only $1$- and $2$-body terms, that is, their Pauli operator expansion
contains only terms of weight at most two, e.g., $X \otimes I^{n-1}, I
\otimes X \otimes Y \otimes I^{n-3}$.  Let ${\cal Q}$ denote the
complementary subspace of $n$-qubit Hamiltonians containing only $3$-
and more-body terms.  Observe that $su(2^n) = {\cal P} + {\cal Q}$,
i.e., any Hamiltonian can be (uniquely) decomposed as a sum $H =
H_P+H_Q$ of one- and two-body terms, $H_P$, and of three- and
more-body terms, $H_Q$.  Overloading notation, we now define maps
${\cal P}$ and ${\cal Q}$ by ${\cal P}(H) \equiv H_P$ and ${\cal Q}(H)
\equiv H_Q$.  We define a right-invariant Riemannian metric which we
call the \emph{standard metric} by:
\begin{eqnarray} \label{eq:standard_metric}
  \langle H, J \rangle \equiv \frac{\tr(H {\cal P}(J))+
    q \, \tr(H{\cal Q}(J))}{2^n}.
\end{eqnarray}
The term $q$ is a \emph{penalty parameter}\footnote{Note that our
  definition and notation for the penalty parameter has changed
  slightly from~\cite{Nielsen06c,Nielsen06d,Nielsen06b}.}, which is
chosen to be sufficiently large ($q > 4^n$ can be shown to work).  The
intuition behind this choice of $q$ can be understood by viewing the
length $\sqrt{\langle H(t),H(t) \rangle}$ as a cost for applying the
Hamiltonian $H(t)$, and so choosing $q$ exponentially large imposes a
very large cost for the direct application of three- and more-qubit
gates.

The metric of Equation~(\ref{eq:standard_metric}) induces a distance
$d(\cdot, \cdot)$ on $SU(2^n)$, as described above.  This distance
function may be related to quantum gate complexity by the following
inequalities, proved in~\cite{Nielsen06b,Nielsen06c}:
\begin{eqnarray} \label{eq:metric_complexity_relationship}
  \frac{b_0 G(U,\epsilon)^{b_1} \epsilon^{b_2}}{n^{b_3}} \leq d(I,U)
  \leq G(U).
\end{eqnarray}
In the first inequality, $G(U,\epsilon)$ is the \emph{approximate gate
  complexity} $G(U,\epsilon)$ of $U$, defined to be the minimal number
of one- and two-qubit gates required to synthesize some $n$-qubit
unitary operation $V$ such that $\|U-V\| \leq \epsilon$, where $\|
\cdot \|$ is the usual matrix norm, and no ancilla qubits are used in
the computation.  The values $b_0,b_1,b_2$ and $b_3$ are positive
constants.  The papers~\cite{Nielsen06c,Nielsen06d} proved this result
for $b_1 = 1/3, b_2 = 2/3$ and $b_3 = 2$; a value for $b_0$ was not
calculated explicitly, and won't be needed in this paper in any case.
Examination of~\cite{Nielsen06c,Nielsen06d} reveals that it is
certainly possible to obtain stronger bounds on these constants, but a
tight analysis remains to be done.  In the second inequality in
Equation~(\ref{eq:metric_complexity_relationship}), $G(U)$ is the
\emph{exact gate complexity} of $U$, i.e., the minimal number of one-
and two-qubit gates required to synthesize $U$ exactly, with no
ancilla qubits used in the synthesis.  Thus, these two inequalities
may be summarized by saying that the distance $d(I,U)$ gives us both a
lower bound on the exact gate complexity, and an upper bound on the
approximate gate complexity of synthesizing $U$.

Equation~(\ref{eq:metric_complexity_relationship}) can be generalized
by using universal families of Hamiltonians other than the one- and
two-qubit Hamiltonians.  The idea is to choose a subspace ${\cal P}'$
of $n$-qubit Hamiltonians which has only polynomial (in $n$)
dimension, and which is universal for computation.  We choose a
subspace ${\cal Q}'$ such that $su(2^n) = {\cal P}'+{\cal Q}'$.
Overloading notation as before, we define a new metric,
\begin{eqnarray} \label{eq:projective_metric}
  \langle H, J \rangle' \equiv \frac{\tr(H {\cal P}'(J))+
    q \, \tr(H{\cal Q}'(J))}{2^n}.
\end{eqnarray}
It can be shown using the techniques of~\cite{Nielsen06c,Nielsen06d}
that provided $q$ is sufficiently large, then
\begin{eqnarray} \label{eq:metric_complexity_relationship_2}
  {\rm poly}(G'(U,\epsilon),n,\epsilon) \leq d'(I,U) \leq G'(U),
\end{eqnarray}
where ${\rm poly}$ is some polynomial (with constant but possibly
non-integer powers), and the primes indicate the change in choice of
universal gates and of metric.  More generally, we shall say that any
metric related to gate complexity by a relationship of the form of
Equation~(\ref{eq:metric_complexity_relationship_2}) is a
\emph{computational metric}.

The metric of most interest to us in the remainder of the paper is the
standard metric of Equation~(\ref{eq:standard_metric}).  However, many
of the results we prove will hold in more generality, for
\emph{projective metrics}, by which we mean a metric of the form of
Equation~(\ref{eq:projective_metric}), or even more generally for
right-invariant metrics.

\section{The connection}
\label{sec:connection}

In this section we derive several explicit formulae for the
Levi-Civita connection, which is the basic geometric object on the
manifold.  These results are standard in the literature on
right-invariant Riemannian manifolds (see, e.g., Appendix~2
in~\cite{Arnold89a} for an introduction), although our derivation is
from a slightly unusual point of view, being based on the background
in differential geometry most common to physicists.  Many of the same
results can be derived from a more abstract point of view by starting
with the general formula for the Levi-Civita connection (e.g.,
Equation~(5.1) on page~69 of~\cite{Lee97a}).  We derive these results
here in part for completeness, and also because it provides an
opportunity to introduce many items of notation and nomenclature used
in subsequent sections.  Note that Appendix~\ref{sec:curvature} uses
these results to compute all the natural curvature quantities
associated to our metric.

If $Y$ and $Z$ are vector fields on a manifold, then in a fixed
co-ordinate system $\{ x^k \}$ the connection is given by:
\begin{eqnarray} \label{eq:connection_coordinate_formula}
  (\nabla_Y Z)^j = \frac{\partial z^j}{\partial x^k} y^k
  + \Gamma^j_{kl} y^kz^l,
\end{eqnarray}
where $y^k, z^k$ are natural co-ordinate representations for the
vector fields $Y$ and $Z$ with respect to the co-ordinate system $\{
x^k \}$, summation over repeated indices is implied, and
$\Gamma^j_{kl}$ are the Christoffel coefficients.  Expressing the
metric tensor $g_{jk}$ with respect to the same system of co-ordinates
we have:
\begin{eqnarray}
  \Gamma^j_{kl} = g^{jm} \Gamma_{mkl} = \frac{g^{jm}}{2}
  \left( g_{m k,l}+g_{m l,k}-g_{kl,m} \right),
\end{eqnarray}
where a subscript $_{,l}$ denotes a partial derivative with respect to
$x_l$.

We have defined our metric in Equation~(\ref{eq:standard_metric})
using a representation for tangent vectors which identifies a
Hamiltonian $H$ with the tangent to $e^{-iHt} U$ at $t=0$.  It is
natural to hope that there is a co-ordinate representation for this
Hamiltonian (e.g., the Pauli expansion coefficients) which can be
identified in a natural way with a set of co-ordinates such as the
$x_j$ above.  Unfortunately, such a representation does not exist.
Instead, to evaluate the Christoffel symbol we must first introduce a
fixed system of co-ordinates on the manifold, which we shall call
Pauli co-ordinates.  We then re-express the metric in terms of these
Pauli co-ordinates, and use this re-expression to evaluate the
Christoffel coefficients.  This can then be used to obtain co-ordinate
independent representations for the connection and other geometric
objects.

\textbf{Pauli co-ordinates:} To evaluate the connection it is
sufficient to introduce co-ordinates defined in a neighbourhood of the
origin, $I \in SU(2^n)$.  This enables us to evaluate the connection
at the origin, and the right-invariance of the metric then gives a
general expression for the connection everywhere on the manifold.  We
define Pauli co-ordinates by representing $U$ in a neighbourhood
of the origin as $U= e^{-iX}$.  Such an $X$ can be defined in a unique
way using the standard branch of the logarithm.  We can associate a
co-ordinate vector $x$ to $X$ via $X = x \cdot \sigma$, i.e., the
vector $x$ consists of the Pauli expansion coefficients of $X$.  Note
that $x^\sigma = \tr(X\sigma)/2^n$.  We call the matrix $X$ the
\emph{Pauli representation} for $U$, and the components $x^\sigma$ the
\emph{Pauli co-ordinates}.  We shall also use this terminology for the
natural corresponding co-ordinates on the tangent spaces, and on the
tangent bundle.

We have defined the metric in terms of a Hamiltonian representation
for tangent vectors.  Our goal now is to re-express the metric in
terms of the Pauli representation. Suppose we are at a point $U =
e^{-iX}$ on the manifold, and that a tangent vector is specified there
by the Hamiltonian $H$.  What is the corresponding representation of
the tangent vector in the Pauli co-ordinate representation for the
tangent space $T_U SU(2^n)$?

To answer this question, we rewrite the tangent curve $e^{-iHt}
e^{-iX}$ corresponding to $H$ in the form $e^{-iHt} e^{-iX} =
e^{-i(X+t J)} + O(t^2)$.  The Baker-Campbell-Hausdorff formula gives
us:
\begin{eqnarray}
  H = {\cal E}_X(J) \equiv
  i \ad_X^{-1} \left( e^{-i \ad_X} - {\cal I} \right)(J),
\end{eqnarray}
where $\ad_X(Y) \equiv [X,Y]$, $[\cdot,\cdot]$ denotes the matrix
commutator, and ${\cal I}(Y) \equiv Y$.  This formula relates the
Hamiltonian representation, $H$, of the tangent vector, to the Pauli
representation, $J$, of the same tangent vector.

It can be shown that ${\cal E}_X$ is invertible near the origin, i.e.,
for $X$ sufficiently close to $0$.  We denote the inverse by ${\cal
  D}_X = {\cal E}_X^{-1}$, so we have $J = {\cal D}_X(H)$.  Note that
${\cal E}_X$ has the power series expansion
\begin{eqnarray}
  {\cal E}_X  =\sum_{j=0}^{\infty} \frac{(-i\rm{ad}_X)^j}{(j+1)!}.
\end{eqnarray}
To compute the connection at the origin it suffices to have expansions
of ${\cal E}_X$ and ${\cal D}_X$ to first order in $X$:
\begin{eqnarray} \label{eq:first_order_expansions}
  {\cal E}_X = {\cal I}-\frac{i\ad_X}{2} + O(X^2);
 \,\,\,\, {\cal D}_X = {\cal I}+\frac{i\ad_X}{2} + O(X^2). \nonumber \\
\end{eqnarray}
Finally, it is helpful to note that the adjoint of these
superoperators with respect to the trace inner product $(X,Y) \equiv
\tr(X^\dagger Y)$ satisfies ${\cal E}_X^\dagger = {\cal E}_{-X}$ and
${\cal D}_X^\dagger = {\cal D}_{-X}$.

\textbf{The metric in Pauli co-ordinates:} In
Section~\ref{sec:background} we defined the metric with respect to the
Hamiltonian representation.  We now rewrite the metric in a
neighbourhood of the origin with respect to the Pauli co-ordinates.
This procedure can be carried out for a general right-invariant metric
with essentially no extra effort beyond what is required for the
standard metric.  The most general form for a right-invariant metric
in the Hamiltonian representation is:
\begin{eqnarray}
  \langle H, J \rangle = \frac{\mbox{tr}(H {\cal G}(J))}{2^n},
\end{eqnarray}
where ${\cal G}$ is a strictly positive (and thus self-adjoint)
superoperator, i.e., a linear operator on matrices such that
$\mbox{tr}(H{\cal G}(H)) > 0$ whenever $H \neq 0$.  For the standard
metric, the explicit form of ${\cal G}$ is ${\cal G} = {\cal P} + q
{\cal Q}$.  It will be convenient for later use to define a Hermitian
matrix $L \equiv {\cal G}(H)$ \emph{dual} to the Hamiltonian, $H$.
Note that the dual satisfies $\langle H, J \rangle = \tr(LJ)/2^n$ for
all $J$.

Suppose $Y$ and $Z$ are tangent vector fields defined in a
neighbourhood of the origin.  We will also use $Y$ and $Z$ to denote
the specific values these vector fields take at a point $U = e^{-iX}$
in that neighbourhood.  Let $Y^P, Z^P$ denote the Pauli representation
for these vectors, and $Y^H = {\cal E}_X(Y^P)$ and $Z^H = {\cal
  E}_X(Z^P)$ denote the corresponding Hamiltonian representation. The
metric then is given by:
\begin{eqnarray}
  \langle Y,Z \rangle & = &
  \frac{\tr(Y^H {\cal G}(Z^H))}{2^n} \\
  & = & \frac{\tr({\cal E}_X(Y^P) {\cal G} \circ {\cal E}_X(Z^P))}{2^n} \\
  & = & \frac{\tr(Y^P {\cal E}_X^\dagger \circ {\cal G} \circ {\cal E}_X(Z^P))}
  {2^n}.
\end{eqnarray}
Define ${\cal G}_X \equiv {\cal E}_X^\dagger \circ {\cal G} \circ
{\cal E}_X$, so $\langle Y,Z\rangle = \tr(Y^P {\cal G}_X(Z^P))/2^n$.
With respect to the Pauli co-ordinates $y^\sigma$ and $z^\sigma$ we
have $Y^P = \sum_\sigma y^\sigma \sigma$ and $Z^P = \sum_\sigma
z^\sigma \sigma$, and thus in this co-ordinate representation the
metric tensor has components
\begin{eqnarray} \label{eq:metric_components}
  g_{\sigma \tau} = \frac{\tr(\sigma {\cal G}_X(\tau))}{2^n}.
\end{eqnarray}
The inverse $g^{\sigma \tau}$ is given by
\begin{eqnarray} \label{eq:inverse_metric_components}
  g^{\sigma \tau} = \frac{\tr(\sigma {\cal F}_X(\tau))}{2^n},
\end{eqnarray}
where ${\cal F}_X \equiv {\cal G}_X^{-1} = {\cal D}_X \circ {\cal F}
\circ {\cal D}_X^\dagger$ and ${\cal F} \equiv {\cal G}^{-1}$.  To
compute first derivatives, we note that
\begin{eqnarray}
  {\cal G}_X = {\cal G} + \frac{i}{2} [\ad_X,{\cal G} ]
   + O(X^2).
\end{eqnarray}
Using the cyclic property of trace, and the fact ${\cal G}^\dagger =
{\cal G}$, a computation shows that at the origin
\begin{eqnarray}
  g_{\sigma\tau,\mu} 
  = \frac{i \, \tr(([{\cal G}(\sigma),\tau]+[{\cal G}(\tau),\sigma])\mu)
    }{2^{n+1}}.
\end{eqnarray}
Other equivalent forms are possible; this form seems to us to be
particularly easy to recall.

\textbf{The Christoffel symbol:} The Christoffel symbol
$\Gamma^\rho_{\sigma \tau} = g^{\rho \mu}\Gamma_{\mu \sigma \tau}$ may
be computed at the origin by observing that
\begin{eqnarray}
  \Gamma_{\mu\sigma \tau} &= & \frac{1}{2} 
\left( g_{\mu\sigma,\tau}+g_{\mu \tau,\sigma}-g_{\sigma \tau,\mu}\right) \\
  & = & \frac{i}{2^{n+1}}
  \tr \left( \mu ([\sigma,{\cal G}(\tau)]+[\tau,{\cal G}(\sigma)]) \right).
\end{eqnarray}
Using $g^{\rho \mu} = \tr({\cal F}(\rho)\mu)/2^n$ we obtain
\begin{eqnarray} \label{eq:Christoffel}
  \Gamma^\rho_{\sigma \tau} = \frac{i}{2^{n+1}} \tr\left(
    {\cal F}(\rho) ([\sigma,{\cal G}(\tau)]+[\tau,{\cal G}(\sigma)]) \right).
\end{eqnarray}
It is worth noting that this formula holds in considerable generality.
The only critical property of the Pauli matrices used in the derivation
is that they are orthonormal (up to a constant factor) with respect to
the trace inner product.

\textbf{The connection:} Working in the Pauli representation and using
Equations~(\ref{eq:connection_coordinate_formula})
and~(\ref{eq:Christoffel}), we have at the origin:
\begin{eqnarray} \label{eq:connection}
  (\nabla_Y Z)^P = y^\sigma Z^P_{,\sigma} + \frac{i}{2} {\cal F}\left(
    [Y^P,{\cal G}(Z^P)]+[Z^P,{\cal G}(Y^P)] \right). \nonumber \\
\end{eqnarray}
This equation gives a formula for the connection evaluated at the
origin, when the vector fields are written in the Pauli
representation.  This can be re-expressed in the Hamiltonian
representation by observing that at the origin $(\nabla_Y Z)^H =
(\nabla_Y Z)^P$, $Y^H = Y^P$ (and thus the $y^\sigma$ components are
the same in both representations), and $Z^H = Z^P$.  Finally, we have
$Z^P = {\cal D}_X(Z^H)$ near the origin, and thus using
Equation~(\ref{eq:first_order_expansions}) we obtain $Z^P_{,\sigma} =
(i/2) [\sigma,Z^H] +Z^H_{,\sigma}$ at the origin.  This gives:
\begin{eqnarray}  \label{eq:connection_Hamiltonian}
  (\nabla_Y Z)^H & = & y^\sigma Z^H_{,\sigma} + \frac{i}{2}\left(
      [Y^H,Z^H] \right. \nonumber \\
    & & \left. + {\cal F}\left(
    [Y^H,{\cal G}(Z^H)]+[Z^H,{\cal G}(Y^H)] \right)\right). \nonumber \\
\end{eqnarray}
Note that the partial derivative in $Z^H_{,\sigma}$ is still with
respect to the Pauli co-ordinates $x^\sigma$ on the manifold.  

Suppose now that we have a curve that passes through the origin and
that has tangent $Y$ at the origin.  Using
Equation~(\ref{eq:connection_Hamiltonian}) we see that the covariant
derivative along the curve $D_t Z \equiv \nabla_Y Z$ is given in the
Hamiltonian representation by:
\begin{eqnarray} \label{eq:Hamiltonian_derivative}
  (D_t Z)^H & = & \frac{dZ^H}{dt}+ \frac{i}{2}\left( [Y^H,Z^H] \right. \nonumber
  \\ 
  & & \left. + {\cal F}\left(
    [Y^H,{\cal G}(Z^H)]+[Z^H,{\cal G}(Y^H)] \right) \right). \nonumber \\
 & &
\end{eqnarray}
Note that because of the right-invariance of the metric this equation
is true everywhere on the manifold.  

The formula for the connection simplifies in the special case when $Z$
is a right-invariant vector field.  In this case $Z^H$ does not vary
as a function of position, and so
Equation~(\ref{eq:connection_Hamiltonian}) gives:
\begin{eqnarray} \label{eq:connection_right_invariant}
  & & (\nabla_Y Z)^H \nonumber \\
  & = & \frac{i}{2} \left( [Y^H,Z^H] + {\cal F}\left(
    [Y^H,{\cal G}(Z^H)]+[Z^H,{\cal G}(Y^H)] \right)\right). \nonumber \\
\end{eqnarray}
Once again, because of the right-invariance of the metric this
equation is true everywhere on the manifold.  A consequence of this
equation that will be useful later is that $\langle X, \nabla_{Y} Z
\rangle = -\langle \nabla_{Y} X, Z \rangle$ for any triple of
right-invariant vector fields, $X, Y$ and $Z$.

Note that in subsequent sections we work almost entirely in the
Hamiltonian representation.  As a consequence, when applying formulae
like Equations~(\ref{eq:connection_Hamiltonian})
and~(\ref{eq:Hamiltonian_derivative}) we will drop the superscript
$H$'s.

\section{Geodesics}
\label{sec:geodesics}

In this section we present the geodesic equation
(Subsection~\ref{subsec:geodesic_equation}), develop a complete set of
constants of the motion for the geodesic equation
(Subsection~\ref{subsec:constants_of_the_motion}), and describe
several classes of analytic solutions to the geodesic equation
(Subsection~\ref{subsec:analytic_solutions}).

\subsection{The geodesic equation}
\label{subsec:geodesic_equation}

By definition, a geodesic in $SU(2^n)$ is a curve $U(t)$ whose tangent
vector $H(t)$ satisfies the condition $D_t H= 0$, i.e., the tangent
vector is parallel transported along the curve.  Using
Equation~(\ref{eq:Hamiltonian_derivative}) this becomes
\begin{eqnarray}
  0 = \dot H + i {\cal F}([H,{\cal G}(H)]).
\end{eqnarray}
This equation is more conveniently rewritten in terms of the dual $L
\equiv {\cal G}(H) = {\cal F}^{-1}(H)$.  After a little algebra we
obtain the geodesic equation in the form we shall most commonly apply
it,
\begin{eqnarray} \label{eq:geodesic_equation}
  \dot L = i[L,{\cal F}(L)].
\end{eqnarray}
We shall refer to this equation as \emph{the} geodesic equation, to
distinguish it from other equivalent forms.  A third form which is
often used in the literature is the form $\langle \dot H, J \rangle =
i \langle H, [H,J]\rangle$, valid for any $J \in su(2^n)$.  Note that
the geodesic equation is a well-known result in the literature on
right-invariant Riemannian manifolds (see, e.g., Appendix~2
of~\cite{Arnold89a}).  Note also that
Equation~(\ref{eq:geodesic_equation}) is in the class of \emph{Lax
  equations} well known to mathematicians.

In the special case of the standard metric the geodesic equation
simplifies nicely.  Recalling that ${\cal G} = {\cal P}+q{\cal Q}$,
and thus ${\cal F} = {\cal G}^{-1} = {\cal P}+q^{-1} {\cal Q}$, we
obtain
\begin{eqnarray} \label{eq:standard_geodesic}
  \dot L & = & i \left(1-q^{-1}\right) [L, {\cal P}(L)].
\end{eqnarray}
Provided $q \neq 1$ we can remove the dependence on $q$ by defining a
rescaled version of $L$, $M \equiv (1-q^{-1}) L$, obtaining a form of
the geodesic equation independent of $q$, except for the requirement
$q \neq 1$:
\begin{eqnarray} \label{eq:rescaled_geodesic_equation}
  \dot M= i[M,{\cal P}(M)].
\end{eqnarray}
This is a remarkable equation.  Because the minimal length path
between any two points on a Riemannian manifold is a geodesic, we may
assume without loss of generality that the optimal Hamiltonian $H(t)$
generating any unitary operation is determined by a solution $M(t)$ to
Equation~(\ref{eq:rescaled_geodesic_equation}) via the rescaling $H =
{\cal G}(M)/(1-q^{-1})$.
Equation~(\ref{eq:rescaled_geodesic_equation}) is thus a single
universal equation whose solutions determine the paths of minimal
length on the manifold.  This situation is in vivid contrast with how
we usually think about the standard circuit model of quantum
computing, where $H(t)$ may have arbitrary time
dependence\footnote{This appears to be a major advantage of the
  geometric approach over the circuit approach.  It comes, however, at
  a cost.  In the geometric approach $H(t)$ may include (exponentially
  small) three- and more-body terms, while in the circuit model only
  one- and two-body terms appear in the Hamiltonian.}.  It is also
notable that Equation~(\ref{eq:standard_geodesic}) is (arguably) the
simplest and most elegant equation involving both the Lie group
structure, expressed through the Lie bracket, and also the map ${\cal
  P}$ onto the set of Hamiltonians that are regarded as
computationally easy to implement.

As a caveat to this optimistic picture, note that being a geodesic is
merely a necessary condition, not a sufficient condition, for a path
to be minimal.  In particular, there may be many geodesics passing
from $I$ to a desired unitary, $U$, and not all those geodesics will
be globally minimizing\footnote{They are, however, locally minimizing
  in the sense that any sufficiently small arc along any geodesic is
  always a global minimum of the length functional (\cite{Berger03a},
  pp~222-226).}.  The situation is analogous to minimizing a function
$f(x)$ in conventional calculus: the condition $f'(x) = 0$ is a
constraint that must be satisfied by $x$ minimizing $f(x)$, but
further analysis is necessary to determine if $f(x)$ is a global
minimum.  Thus finding geodesics is only a first step towards the
determination of the distance $d(I,U)$.

\subsection{Constants of the motion}
\label{subsec:constants_of_the_motion}

The geodesic equation (Equation~(\ref{eq:geodesic_equation})) on a
right-invariant Riemannian manifold has a corresponding set of
constants of the motion which completely determine the
geodesics~\cite{Arnold89a}.  To see this, observe that for any choice
of $L_0$ the function $L(t) = U(t) L_0 U(t)^\dagger$ satisfies the
geodesic equation, Equation~(\ref{eq:geodesic_equation}).  It follows
that along any geodesic the function $U(t)^\dagger L(t) U(t) = L_0$ is
a matrix-valued constant of the motion.  Furthermore, by
differentiating the equation $L(t) = U(t) L_0 U(t)^\dagger$ we may
recover the geodesic equation, and thus this set of constants of the
motion completely determines the geodesics of the system.

As an aside, it is possible to \emph{derive} these constants of the
motion (and thus the geodesic equation) by observing that the metric
is invariant under a continuous symmetry group, namely, arbitrary
right translations of $SU(2^n)$. One may then use Noether's theorem to
find the associated constants of the motion, which turn out to be
precisely the matrix elements of $U(t)^\dagger L(t) U(t)$.  This is
the approach taken to derive the geodesic equation
in~\cite{Arnold89a}.

\textbf{One-body terms are constants of the motion:} In the special
case of the standard metric, it can be shown that the coefficients of
the one-body terms in the Pauli expansion of $H(t)$ are also constants
of the motion along geodesics.  This fact is useful in developing
certain analytic solutions to the geodesic equation, to be described
later.  To prove that the one-body terms are constant, let ${\cal
  S}(X)$ map the $n$-qubit matrix $X$ onto just its one-body terms.
We see that:
\begin{eqnarray}
  \frac{d{\cal S}(L)}{dt} & = & {\cal S}\left(\frac{dL}{dt}\right) \\
  & = & i (1-q^{-1}){\cal S}([L,{\cal P}(L)])\\
  & = & i (1-q^{-1}){\cal S}([{\cal Q}(L),{\cal P}(L)]).
\end{eqnarray}
The Pauli commutation relations imply that: (1) the commutator of
${\cal Q}(L)$ with the one-body terms in ${\cal P}(L)$ produces
only three- and more-body terms, and (2) the commutator of ${\cal
  Q}(L)$ with the two-body terms in ${\cal P}(L)$ produces only two-
and more-body terms.  As a result ${\cal S}$ annihilates $[{\cal
  Q}(L),{\cal P}(L)]$, and so ${\cal S}(L)$ is a constant of the
motion.

An alternative proof that the one-body terms are constants of the
motion may be found by applying Noether's theorem to the continuous
symmetry $SU(2^n) \rightarrow SU(2^n)$ defined by $U \rightarrow V U
V^\dagger$, where $V$ is an arbitrary one-qubit unitary.

\subsection{Analytic solutions to the geodesic equation}
\label{subsec:analytic_solutions}

We now develop a range of partial and full solutions to the geodesic
equation.  Our results are mostly specialized to projective metrics,
and some results are further specialized to the standard metric.
In~\ref{subsubsec:constant_Hamiltonian} we develop general necessary
and sufficient conditions for geodesics to be of the form $e^{-iHt}$
for a constant Hamiltonian $H$.
In~\ref{subsubsec:three_qubit_geodesics} we find an exact form for the
geodesics of the standard metric in the three-qubit case, for the $q
\rightarrow \infty$ limit.  This form is based on an algebraic
structure that is also useful in other contexts.  Finally,
in~\ref{subsubsec:formal_solution} we develop a formal power series
solution to the geodesic equation.

\subsubsection{Geodesics where the Hamiltonian is constant} 
\label{subsubsec:constant_Hamiltonian}

Along certain geodesics the Hamiltonian is constant, and thus the
geodesic has the form $e^{-iHt}$.  To determine when this is the case
it suffices to determine when the dual $L(t)$ is constant along a
geodesic, since the dual is related to the Hamiltonian by a fixed
invertible transformation.  From the geodesic equation for projective
metrics, $\dot L = i (1-q^{-1})[L,{\cal P}(L)]$, we see that in the $q
= 1$ case $L$ is always constant along geodesics.  However, the case
of most interest to us is when $q$ is very large, where we see that a
necessary and sufficient condition for $L$ to be constant is that
$[{\cal Q}(L),{\cal P}(L)]= 0$.  This is equivalent to the condition
that $[{\cal Q}(H),{\cal P}(H)] = 0$.  When this condition is
satisfied, and only when it is satisfied, the geodesic is of the form
$\exp(-iHt)$.

In the case of the standard metric, we see that this condition is that
the one- and two-body terms in a Hamiltonian should commute with the
three- and more-body terms.  An appealing consequence is that whenever
$H$ contains only one- and two-body terms, then $\exp(-iHt)$ is a
geodesic.  Over sufficiently short time periods geodesics are
guaranteed to be minimal length curves (see, e.g., Section~3.3
of~\cite{doCarmo92a}), and so this result accords with the intuition
that the fastest way to simulate a physical system is with its own
evolution.

\subsubsection{Geodesics for three qubits} 
\label{subsubsec:three_qubit_geodesics}

In the case where there are only three qubits we can derive a solution
to the geodesic equation for the standard metric that is exact in the
large $q$ limit.  In fact, it turns out to be possible to analyze a
more general class of metrics, defined by the choice
\begin{eqnarray}
  {\cal G} \equiv s {\cal S} + {\cal T} + q {\cal Q}, 
\end{eqnarray}
where ${\cal S}$ maps onto the subspace of three-qubit Hamiltonians
which contain only one-body terms, ${\cal T}$ maps onto the subspace
containing only two-body terms, and ${\cal Q}$ maps onto the subspace
containing only three-body terms.  In the case $s = 1$ this reduces to
the standard metric.  The limit $s \rightarrow 0$ corresponds to the
case where one-body Hamiltonians may be applied effectively for free.

The key observation needed to obtain the geodesics is the commutation
relations between the matrix subspaces ${\cal S}$, ${\cal T}$ and
${\cal Q}$,
\begin{eqnarray} \label{eq:comm_relations_three_qubit}
  [{\cal S},{\cal T}] & \subseteq & {\cal T} \\
  {} [{\cal S},{\cal Q}] & \subseteq & {\cal Q} \\
  {} [{\cal T},{\cal Q}] & \subseteq & {\cal T}. \label{eq:comm_relations_three_qubit_terminating}
\end{eqnarray}
Note that the derivation which follows depends only on these
commutation relations, and not on the specific choice of ${\cal S}$ as
one-body Hamiltonians, etcetera.  It would be interesting to obtain
other examples, outside the three-qubit context, where this algebraic
structure appears naturally.  We do not know whether this particular
structure is ever of computational interest in the large $n$ limit.
We note that this general approach of using algebraic structure to
obtain insight into geometry is reminiscent of the work of Khaneja,
Glaser and Brockett~\cite{Khaneja01a}, who make use of symmetric
spaces to solve geometric problems involving two qubits.

Defining $S \equiv {\cal S}(L), T \equiv {\cal T}(L)$, and $Q \equiv
{\cal Q}(L)$, we see from the commutation relations of
Equations~(\ref{eq:comm_relations_three_qubit})-(\ref{eq:comm_relations_three_qubit_terminating})
that the geodesic equation $\dot L = i [L,{\cal F}(L)]$ becomes:
\begin{eqnarray}
  \dot S & = & 0 \\
  \dot T & = & i \left[ (1-s^{-1}) S + (1-q^{-1})Q,T \right] \\
  \dot Q & = & i(q^{-1}-s^{-1}) [S,Q].
\end{eqnarray}
To solve these equations, observe that $S$ is a constant of the
motion.  This makes the equation for $Q$ a linear equation that is
easily solved.  The equation for $T$ is then a time-dependent linear
equation that can be solved using standard techniques.  The resulting
solution is
\begin{eqnarray}
  S(t) & = & S_0 \\
  T(t) & = & e^{it(q^{-1}-s^{-1})S_0} e^{it(1-q^{-1})(S_0+Q_0)}\times
  \nonumber  \\
 & & T_0 e^{-it(1-q^{-1})(S_0+Q_0)} e^{-it(q^{-1}-s^{-1})S_0} \\
  Q(t) & = & e^{it(q^{-1}-s^{-1})S_0}Q_0  e^{-it(q^{-1}-s^{-1})S_0}.
\end{eqnarray}
The corresponding Hamiltonian has the form:
\begin{eqnarray}
  H(t) = s^{-1} S(t) + T(t) + q^{-1} Q(t).
\end{eqnarray}
This expression for the Hamiltonian holds for all $q$ and $s$.  We
now show how to integrate the corresponding Schr\"odinger
equation in the large $q$ limit to obtain the geodesic $U(t)$.

Without loss of generality we can assume that we are working on a
geodesic with $\langle H(t),H(t) \rangle = 1$ for all time. As a
result we obtain the bounds $\mbox{tr}(S^2)/2^3 \leq s,
\mbox{tr}(T^2)/2^3 \leq 1$, and $\mbox{tr}(Q^2)/2^3 \leq q$.  The term
$q^{-1}Q(t)$ is therefore of order $q^{-1/2}$, and thus may be
neglected in the large $q$ limit, with a resulting error in $U(t)$ of
order $t q^{-1/2}$.  For similar reasons, we can neglect the $q^{-1}$
terms in the exponentials appearing in $T(t)$.  The resulting error in
$T(t)$ is at most of order $t (s^{1/2}q^{-1}+q^{-1/2})$, and thus the
error in $U(t)$ is at most of order $t^2 (s^{1/2} q^{-1}+q^{-1/2})$.
This leads us to define an approximate Hamiltonian
\begin{eqnarray}
  \tilde H(t) & = & s^{-1} S_0 +e^{-it s^{-1}S_0} e^{it(S_0+Q_0)}\times
  \nonumber  \\
 & & T_0 e^{-it(S_0+Q_0)} e^{its^{-1}S_0}.
\end{eqnarray}
The corresponding solution $\tilde U(t)$ to the Schr\"odinger equation
satisfies
\begin{eqnarray}
  \|U(t) - \tilde U(t) \| \leq O(t q^{-1/2}+t^2(s^{1/2}q^{-1}+q^{-1/2})).
\end{eqnarray}
Making the change of variables $\tilde V = e^{-it(S_0+Q_0)}e^{i
  ts^{-1} S_0} \tilde U$ we see that the Schr\"odinger equation is
equivalent to
\begin{eqnarray}
  \dot{\tilde V} = -i \left( S_0+T_0+Q_0 \right) \tilde V.
\end{eqnarray}
The approximate solution to the geodesic equation is thus
\begin{eqnarray} \label{eq:algebraic_geodesic_solution}
  \tilde U(t) = e^{-it s^{-1} S_0} e^{it(S_0+Q_0)}e^{-it(S_0+T_0+Q_0)}.
\end{eqnarray}
Although this form is an exact solution to the geodesic equation in
the $q \rightarrow \infty$ limit, it is not obvious which is the
minimal geodesic passing through a particular desired unitary $U$.
Developing techniques to find minimal geodesics in this case is an
interesting problem for further work.

The special case $s \rightarrow 0$ is of some interest in our
three-qubit example, where it corresponds to zero cost for local
unitary operations.  In this limit the $S_0$ terms in the second and
third exponentials of Equation~(\ref{eq:algebraic_geodesic_solution})
may be neglected, and we obtain the solution:
\begin{eqnarray}
  \tilde U(t) = e^{-it s^{-1} S_0} e^{itQ_0}e^{-it(T_0+Q_0)}.
\end{eqnarray}

Returning to the case of general $s$, we now attempt to simplify the
expression in Equation~(\ref{eq:algebraic_geodesic_solution}).
Generically, we expect that $S_0+Q_0$ is large compared with $T_0$,
and $S_0+Q_0$ is non-degenerate.  First-order perturbation theory can
be used to simplify the product of the final two terms to obtain
\begin{eqnarray}
  \tilde U(t) = e^{-it s^{-1} S_0} e^{-it {\cal R}_{S_0+Q_0}(T_0)},
\end{eqnarray}
where ${\cal R}_{S_0+Q_0}(T_0)$ denotes the diagonal matrix which
remains when we work in the eigenbasis of $S_0+Q_0$ and remove all
off-diagonal entries from $T_0$.  Assuming $Q_0$ is nondegenerate, in
the $s \rightarrow 0$ limit we obtain
\begin{eqnarray} \label{eq:simplified_form_geodesic}
  \tilde U(t) = e^{-it s^{-1} S_0} e^{-it {\cal R}_{Q_0}(T_0)}.
\end{eqnarray}

\subsubsection{Formal solution of the geodesic equation} 
\label{subsubsec:formal_solution}

In this section we develop a formal power series solution to the
geodesic equation.  The formal solution is most easily developed for
the dual $L(t)$ which satisfies the equation $\dot L = i[L,{\cal
  F}(L)]$.  We expand $L(t)$ in a power series,
\begin{eqnarray}
  L(t) = \sum_{j=0}^\infty \frac{L^{(j)}(0)t^j}{j!},
\end{eqnarray}
where $L^{(j)}(0)$ is the $j$'th derivative of $L$(t) at $t=0$.  This
derivative can, in principle, be evaluated using the geodesic
equation.  It is rather inconvenient to do this directly.  However, it
can be done easily using the vectorization technique, whereby matrices
are converted into vectors, and linear operations taking matrices to
matrices become matrix operations taking vectors to vectors.  We
assume readers are familiar with vectorization (see, e.g., Chapter~4
of~\cite{Horn91a}).

We will write the formal solution for any equation of the form $\dot L
= {\cal E}(L,L)$, where ${\cal E}(\cdot,\cdot)$ is a bilinear
operation.  This class of equations includes the geodesic equation for
any right-invariant metric.  Vectorization of this equation yields
\begin{eqnarray}
  \frac{d|L)}{dt} = E (|L)\otimes|L)),
\end{eqnarray}
where $|L)$ is the vectorized form of the matrix $L$, and $E$ is the
vectorized form of the bilinear operation ${\cal E}$.  Note that $E$
is a linear operation mapping from the tensor product of two copies of
the space on which $|L)$ lives into a single copy of that space.  For
the class of operations ${\cal E}$ arising from the geodesic equation,
standard vectorization techniques show that $E$ has the explicit form
$E = i R(I-S) (I\otimes F)$, where $F$ is the vectorized form of
${\cal F}$, $S$ swaps the factors in the tensor product, and $R$ is
defined by $R(|X) \otimes |Y)) \equiv |XY)$, where $XY$ is the usual
matrix product of $X$ and $Y$.

Taking repeated derivatives, it follows that the $j$'th derivative may
be written
\begin{eqnarray} \label{eq:geodesic_derivative}
  |L^{(j)}) & = & E(E \otimes I+I \otimes E) \ldots \nonumber \\
  & & (E \otimes I^{\otimes(j-1)} + I \otimes
  E \otimes I^{\otimes(j-2)} + \ldots) \nonumber \\
  & & |L)^{\otimes (j+1)},
\end{eqnarray}
where it is understood that $|L^{(j)})$ and $|L)$ are evaluated at
time $t=0$.  To simplify this expression, observe that:
\begin{eqnarray}
  I^{\otimes k} \otimes E \otimes I^{\otimes l} =
  S_{1,k+1} (E \otimes I^{\otimes (k+l)}) \pi,
\end{eqnarray}
where $S_{1,k}$ swaps systems $1$ and $k$, and acts trivially on all
other systems, and $\pi$ is some permutation of the systems.  It
follows that if $|X)$ is a vector in the entire tensor product space
such that $|X)$ is symmetric under interchange of any of the systems,
then we have
\begin{eqnarray}
  (I^{\otimes k} \otimes E \otimes I^{\otimes l})|X) =
  S_{1,k+1} (E \otimes I^{\otimes (k+l)})|X),
\end{eqnarray}
where we used the fact that $\pi|X) = |X)$ for all permutations $\pi$.
Define an operator $T_m$ acting on $m$ systems by
\begin{eqnarray}
  T_m \equiv I+ S_{1,2} + \ldots + S_{1,m},
\end{eqnarray}
where it is understood that each swap $S_{1,j}$ acts on $m$ systems.
Our earlier expression for $|L^{(j)})$,
Equation~(\ref{eq:geodesic_derivative}), may now be rewritten as
\begin{eqnarray}
  |L^{(j)}) & = & E T_2 (E \otimes I) T_3 (E \otimes I^{\otimes 2})
  \nonumber \\
  & & \ldots T_j (E \otimes I^{\otimes(j-1)}) |L)^{\otimes (j+1)}. 
\end{eqnarray}
Thus we have the desired formal expression for the vectorized solution
$|L(t))$ to the equation $\dot L = {\cal E}(L,L)$,
\begin{eqnarray} \label{eq:formal_geodesic_solution}
  |L(t)) = \sum_{j=0}^\infty \frac{\prod_{k=1}^j T_k (E \otimes I^{k-1})
  |L(0))^{\otimes (j+1)}t^j}{j!}.
\end{eqnarray}

\section{Geodesic deformation and conjugate points}
\label{sec:geodesic_deformation}

A central problem in developing our geometric approach to quantum
computation is to find a minimal geodesic from the identity $I$ to a
specified unitary $U$.  In this section we develop two sets of tools
that can be used to make progress towards the solution of this
problem, and illustrate these tools through numerical examples.

The first set of tools are known as Jacobi fields and conjugate
points.  They are standard tools in Riemannian geometry, and relate
the global problem of determining when a geodesic is minimizing to
local curvature properties of the manifold.  We will use these tools
to give explicit examples of geodesics which are provably \emph{not}
minimizing, i.e., they can be used to find examples of curves which
are local length minima, but which are not global length minima.  For
example, for geodesics of the form $\exp(-iHt)$, with $H$ containing
only one- and two-body terms, we use conjugate points to find values
of $t$ beyond which these geodesics are not minimal.

The second set of tools is aimed at solving the geodesic equation with
fixed endpoints $I$ and $U$.  Traditional methods for solving this
two-point boundary value problem (e.g., shooting methods) do not work
so well, since the space we are working in has extremely high
dimensionality.  The idea we use is to \emph{deform} the geodesics
from the $q=1$ case, where the form of the geodesics is well
understood, to much larger values of the penalty, e.g. $q=4^n$.  We
will show that this deformation can be achieved using a generalization
of the Jacobi equation, which we call the \emph{lifted} Jacobi
equation.  The lifted Jacobi equation enables us to define a notion of
\emph{geodesic derivative}, which is a way of deforming the geodesic
as the penalty $q$ is varied, without changing the endpoints.  The
lifted Jacobi equation, the geodesic derivative, and the deformation
algorithm are all original, so far as we are aware.

The detailed structure of the section is as follows.  We begin in
Subsection~\ref{subsec:lifted_Jacobi_equation} by deriving the lifted
Jacobi equation, and obtain as a special case the standard Jacobi
equation.  This is done on a general Riemannian manifold. In
Subsection~\ref{subsec:specific_lifted_Jacobi_equation} we give
explicit forms of these equations which are applicable to the standard
metric.  In particular, the lifted Jacobi equation describes how
geodesics deform as the parameter $q$ is varied in the standard
metric.  Subsection~\ref{subsec:conjugate_points} uses the
conventional Jacobi equation to numerically investigate conjugate
points, and to find examples of geodesics which are provably not
minimizing.  Subsection~\ref{subsec:geodesic_derivative} defines the
geodesic derivative, and studies its basic properties, including
obtaining necessary and sufficient conditions for the geodesic
derivative to exist.  The geodesic derivative is then applied in
Subsection~\ref{subsec:numerical_geodesics} to obtain a numerical
procedure for finding geodesics between $I$ and a specified goal
unitary, $U$.  We illustrate this procedure with some numerical
examples.

\subsection{Lifted Jacobi equation}
\label{subsec:lifted_Jacobi_equation}

Suppose $\gamma(t)$ is a geodesic on a smooth manifold, $M$, with
respect to some metric, $g$, and we smoothly change that metric.
Intuitively, it seems it should be possible to smoothly deform the
geodesic curve so that it remains a geodesic with respect to the new
metric.  The lifted Jacobi equation provides a way of making this
intuition rigorous.  It generalizes a well-known tool of Riemannian
geometry known as the Jacobi equation, which describes the behaviour
of nearby geodesics of a fixed metric.

We develop the lifted Jacobi equation on a general Riemannian
manifold, $M$, and specialize later to cases of interest in the
context of quantum computing.  We suppose $g_s$ is a family of metric
tensor fields for $M$ parameterized by a single real parameter, $s$,
and smooth with respect to any fixed co-ordinate system.  Define a ${0
  \choose 2}$ symmetric tensor field $g'$ to be the pointwise
derivative of $g$ with respect to $s$ at some fixed value of $s$, say
$s=0$.

Our strategy is as follows.  Imagine $\gamma(s,t)$ is a smooth family
of curves on $M$ such that $\gamma(s,\cdot)$ is a geodesic with
respect to the metric $g_s$.  We call $\gamma(0,\cdot)$ the \emph{base
  geodesic}, and define the \emph{lifted Jacobi field} $J(t) \in
T_{\gamma(0,t)} M$ along the base geodesic by
\begin{eqnarray}
  J(t) \equiv \partial_s \gamma(0,t).
\end{eqnarray}
The lifted Jacobi field is the vector field telling us how the base
geodesic is locally deformed as $s$ is varied.  We will show as a
consequence of the geodesic property that $J(t)$ satisfies the lifted
Jacobi equation, which is a second order differential equation.
Conversely, given a solution to the lifted Jacobi equation, it is
possible to define a corresponding family of deformed geodesics.

To derive the lifted Jacobi equation we expand $\gamma(\Delta,t)$ in a
co-ordinate representation as
\begin{eqnarray}
  \gamma(\Delta,t) = \gamma(0,t) + \Delta J(t) + O(\Delta^2).
\end{eqnarray}
By definition $\gamma(\Delta,t)$ satisfies the geodesic equation
associated to the metric $g_\Delta$.  Substituting into the geodesic
equation, expanding in powers of $\Delta$, and considering the term
linear in $\Delta$ gives
\begin{eqnarray} \label{eq:lifted_Jacobi_intermediate}
  0 & = & \frac{\partial J^j}{\partial t^2}
  + \Gamma^j_{kl} \frac{\partial J^k}{\partial t} \frac{\partial \gamma^l}
  {\partial t}
  + \Gamma^j_{kl} \frac{\partial \gamma^k}{\partial t} \frac{\partial J^l}
  {\partial t}  \nonumber \\
  & & +\Gamma^j_{kl,m} J^m \frac{\partial \gamma^k}{\partial t}
  \frac{\partial \gamma^l}{\partial t}
  +\frac{\partial \Gamma^j_{kl}}{\partial s}
  \frac{\partial \gamma^k}{\partial t} \frac{\partial \gamma^l}{\partial t}.
\end{eqnarray}
The standard Jacobi equation corresponds to the case where $g_s$ is
constant, i.e., to the case where the first four terms on the
right-hand side of the above equation sum to zero.  This allows us to
rewrite the above equation in more geometric terms as
\begin{eqnarray} \label{eq:lifted_Jacobi_intermediate_2}
  (D_t^2 J)^j + (R(J,\dot \gamma) \dot \gamma)^j 
  +\frac{\partial \Gamma^j_{kl}}{\partial s}
  \dot \gamma^k \dot \gamma^l
  = 0.
\end{eqnarray}
The first two terms here are just the standard terms appearing in the
conventional Jacobi equation, with $\dot \gamma(t) \equiv \partial
\gamma(0,t)/\partial t$ and $J(t) \equiv J(0,t)$.  Note that $R$ here
is a ${1 \choose 3}$ tensor field formed by raising the last index of
the Riemann curvature tensor, and thus has components $R_{j k
  l}^{\phantom{jkl} m}$.  A lengthy and tedious but essentially
straightforward calculation can be used to verify that these first two
terms correspond to the first four terms in
Equation~(\ref{eq:lifted_Jacobi_intermediate}).

Equation~(\ref{eq:lifted_Jacobi_intermediate_2}) can be rewritten in a
still more natural geometric form.  A calculation shows that
\begin{eqnarray}
  \frac{\partial \Gamma^j_{kl}}{\partial s}
  = \frac{g^{jm}}{2}\left( g'_{mk;l} + g'_{ml;k}-g'_{kl;m} \right),
\end{eqnarray}
where $g'_{mk;l}$ is the standard notation for the covariant
derivative of the tensor field $g'$.  We see from this equation that
$\partial \Gamma^j_{kl}/\partial s$ is a ${1 \choose 2}$ tensor field.
This is a remarkable fact, given that $\Gamma^j_{kl}$ is not a tensor
field.  (A simple alternate proof that $\partial
\Gamma^j_{kl}/\partial s$ is a tensor field may be obtained by taking
the partial derivative with respect to $s$ of the standard
(non-tensorial) transformation law for $\Gamma^j_{kl})$.

Putting it all together, we obtain the lifted Jacobi equation
\begin{eqnarray} \label{eq:lifted_Jacobi}
  (D_t^2 J)^j + (R(J,\dot \gamma) \dot \gamma)^j 
  +  C^j = 0,
\end{eqnarray}
where
\begin{eqnarray} \label{eq:def_inhomog}
  C^j = \frac{g^{jm}}{2}\left( g'_{mk;l} + g'_{ml;k}-g'_{kl;m} \right)
  \dot \gamma^k \dot \gamma^l
\end{eqnarray}
is a vector field that does not depend on the lifted Jacobi field
$J^j$.  Note that the terms $g'_{mk;l}\dot \gamma^k \dot \gamma^l$ and
$g'_{ml;k} \dot \gamma^k \dot \gamma^l$ appearing in $C^j$ are equal,
which may be used to simplify the form of $C^j$.

We have shown that given a family of curves $\gamma(s,t)$ such that
$\gamma(s,\cdot)$ is a geodesic of the metric $g_s$, the corresponding
lifted Jacobi field $J(t)$ must satisfy the lifted Jacobi equation,
Equation~(\ref{eq:lifted_Jacobi}).  It is straightforward to turn this
reasoning around, and argue that for any solution $J(t)$ to the lifted
Jacobi equation there must exist a family $\gamma(s,t)$ of geodesics
for $g_s$ with $J(t)$ as the corresponding lifted Jacobi field.

\textbf{Solution to the lifted Jacobi equation:} The lifted Jacobi
equation is a linear, inhomogeneous second order differential
equation, and thus it is possible to write a solution to the equation
in terms of time-ordered integrals along the geodesic.  In
co-ordinates the lifted Jacobi equation,
Equation~(\ref{eq:lifted_Jacobi}), may be written as
\begin{eqnarray}
  \frac{d^2J}{dt^2} + A \frac{dJ}{dt} + B J + C = 0,
\end{eqnarray}
where $A$ and $B$ are time-dependent matrices, and $C$ is a
time-dependent vector.  This may be rewritten as a first-order system
by setting $J_1 = J, J_2 = \dot J$, and $K = \left[{J_1 \atop
    J_2}\right]$, so
\begin{eqnarray}
  \frac{dK}{dt} 
  = \left[ \begin{array}{cc} 0 & I \\ -B & -A \end{array} \right]K
  -\left[\begin{array}{c} 0 \\ C \end{array} \right].
\end{eqnarray}
Let $E_t$ denote the propagator describing the solution to this
equation in the homogeneous case, i.e., when $C = 0$ we have $K(t) =
E_t K(0)$.  This corresponds to the solution of the conventional
Jacobi equation.  Note that $E_t$ is a time-ordered exponential which
may be studied using standard techniques.  The solution in the
inhomogeneous case is then
\begin{eqnarray} \label{eq:solution_lifted_Jacobi_equation}
  K(t) = E_t K(0) - E_t \int_0^t dr \, E_r^{-1} 
  \left[\begin{array}{c} 0 \\ C(r) \end{array} \right].
\end{eqnarray}
This expression shows that the solutions to the lifted Jacobi equation
may be obtained from the propagator $E_t$ for the conventional Jacobi
equation, and an integral involving an expression $C(\cdot)$
determined by $g'$.  An interesting special case of the solution
arises when we pick $K(0) = 0$, which corresponds to keeping the
initial position and tangent vector to the geodesic unchanged, and
looking to see how the geodesic deforms.  We obtain in this case
\begin{eqnarray}
  K(t) = - E_t \int_0^t dr \, E_r^{-1} 
  \left[\begin{array}{c} 0 \\ C(r) \end{array} \right].
\end{eqnarray}

\subsection{Lifted Jacobi equation for varying penalty}
\label{subsec:specific_lifted_Jacobi_equation}

In the last section we derived the lifted Jacobi equation for a
general parameterized family of metrics on a Riemannian manifold, $M$.
In this section we derive and present a formal solution to the lifted
Jacobi equation for a parameterized family of right-invariant metrics
such as arise in the context of quantum computation.  Specifically, we
choose the parameterized family of metrics $\mathcal{G}_q =
\mathcal{P}+q \mathcal{Q}$, so that $\mathcal{G}' = \mathcal{Q}$.  We
are able to obtain an explicit solution to the corresponding lifted
Jacobi equation along geodesics for which the Hamiltonian $H$ is
constant.

One way of approaching this task is to begin with the lifted Jacobi
equation in the form derived in the last section,
Equation~(\ref{eq:lifted_Jacobi}).  In fact, for right-invariant
metrics on $SU(2^n)$ there is a simpler alternate approach.  We
suppose $H(t)$ is a Hamiltonian generating a geodesic $U(t)$ for the
metric ${\cal G}_q$, and that there is a nearby ${\cal G}_{q+\Delta}$
geodesic of the form
\begin{eqnarray}
  \tilde U(t) = U(t) e^{-i \Delta J(t)},
\end{eqnarray}
for some small $\Delta$.  We will write the lifted Jacobi equation as
a second order differential equation for $J(t)$.  To derive this
equation, rather than start from Equation~(\ref{eq:lifted_Jacobi}),
which requires converting $J(t)$ into a suitable co-ordinate
representation, and then computing all the relevant quantities, it is
easiest to rework through the strategy in the last section, but
working directly in terms of the quantity $J(t)$ rather than some
co-ordinate representation.

To do this, we use Schr\"odinger's equation to deduce that the
Hamiltonian generating $\tilde U(t)$ is, to first order in $\Delta$,
\begin{eqnarray}
  \tilde H(t) = H(t) + \Delta U(t) \dot J(t) U^\dagger(t) + O(\Delta^2).
\end{eqnarray}
We require that $\tilde H(t)$ satisfies the geodesic equation for
${\cal G}_{q+\Delta}$.  To see what this implies we set $\tilde L =
\tilde {\cal G}(\tilde H)$, where $\tilde {\cal G} = {\cal G} + \Delta
{\cal G}'+O(\Delta^2)$.  Substituting into the geodesic equation
$\dot{\tilde L} = i [\tilde L,\tilde H]$, and examining the terms
linear in $\Delta$, we obtain
\begin{eqnarray}
  0 & = & \dot K  + {\cal F} \left(i[K,L] + i[H,{\cal G}(K)] \right. \nonumber \\
 && \left. + ({\cal G}' \circ {\cal F})(i[L,H]) + i[H,{\cal G}'(H)]) \right) \, , \label{eq:lifted_jacobi_right_invariant}
\end{eqnarray}
where $K = U \dot J U^\dagger$ is the first-order perturbation to the
Hamiltonian. This equation is an inhomogeneous first order
differential equation linear in $K$ and thus can be integrated using
standard techniques, and then integrated again to obtain $J(t)$.  The
conventional Jacobi equation corresponds to the case where ${\cal G}'
= 0$, and thus the last two terms vanish.  Note that we have used no
special features of the standard metric in our derivation, and this
form of the lifted Jacobi equation holds for any right-invariant
metric.

\textbf{The case of constant $H$:} Along geodesics where $H$ is
constant it is possible to obtain a closed form expression for the
solutions to the lifted Jacobi equation corresponding to the standard
metric, i.e., with ${\cal G}' = {\cal Q}$.

To see this, we observe first that the solution to the Jacobi and
lifted Jacobi equations coincide when $H$ is a constant.  This is
because the inhomogeneous contribution $({\cal F} \circ {\cal G}'
\circ {\cal F})(i[L,H])+ {\cal F}(i[H,{\cal G}'(H)])$ to the lifted
Jacobi equation vanishes, since $[L,H] = 0$ and $[H,{\cal G}'(H)] =
[H,{\cal Q}(H)] = 0$ along a geodesic with constant $H$.  An
interesting consequence is that if we choose $J(0) = 0$ and $\dot J(0)
= 0$, then $J(t) = 0$ for all time, i.e., the geodesic does not deform
as the metric is varied.  In other words, the geodesics for which $H$
is constant are the same for all $q$, as can also be seen from the
condition derived in Section~\ref{subsec:geodesic_equation}.

We may thus vectorize the lifted Jacobi equation, obtaining $|\dot K)
= i A |K)$, with
\begin{eqnarray}
 A =  F\left[(I \otimes L -L^T \otimes I)+(H^T \otimes I - I \otimes H)G\right],
  \nonumber \\
\end{eqnarray}
where $G$ and $F$ are the vectorized forms of ${\cal G}$ and ${\cal
  F}$, respectively.  The solution to this equation is $|K(t)) = e^{i
  A t} |K(0))$.  Vectorizing $\dot J = U^\dagger K U$ and substituting
$U = e^{-iHt}$ gives
\begin{eqnarray}
  |\dot J(t)) = e^{i B t} e^{i A t} |\dot J(0)).
\end{eqnarray}
where $B = I \otimes H - H^T \otimes I$.  Integrating we obtain
\begin{eqnarray}
  |J(t)) = |J(0)) + \int_0^t dr \, e^{i B r}e^{i A r} |\dot J(0)).
\end{eqnarray}
This integral can be performed explicitly by using a second level of
vectorization, this time acting on matrices in the space in which $A$ and
$B$ live.  We denote this vectorization operation using
$\mbox{vec}$, to distinguish it from the map $X \rightarrow |X)$, and
use $\mbox{unvec}$ to denote the inverse operation.  The integral can
now be evaluated to yield the explicit solution to the lifted Jacobi
equation,
\begin{eqnarray} \label{eq:explicit_solution_lifted_Jacobi_constant_Ham}
  |J(t)) & = & |J(0)) \\
  & & + \mbox{unvec} \left[
    \frac{e^{i(A^T \otimes I + I \otimes B)t}-I}{i(A^T \otimes I + I \otimes B)}
    \mbox{vec}(I) \right] |\dot J(0)).  \nonumber
\end{eqnarray}
This expression looks daunting, due to the multiple layers of
vectorization, but is actually quite simple.

\subsection{Conjugate points}
\label{subsec:conjugate_points}

In this section we use the theory of conjugate points to derive
conditions under which geodesics are no longer minimizing.  In
particular, we numerically study geodesics of the form $e^{-iHt}$,
where $H$ is a fixed two-body Hamiltonian, and use conjugate points to
derive conditions on $t$ such that the geodesic from $I$ to $e^{-iHt}$
is only a local minimum of the length, not a global minimum.  This
work will also be useful in our later discussion of the geodesic
derivative.

Recall the definition of conjugate points from elementary Riemannian
geometry~\cite{Lee97a}.  Two points $x$ and $y$ along a geodesic are
said to be conjugate if there exists a non-zero Jacobi field defined
along the geodesic which vanishes at both $x$ and $y$.  If we write
the propagator for the Jacobi equation in block form as
\begin{eqnarray}
  E_t = \left[ \begin{array}{cc} E_1 & E_2 \\ E_3 & E_4 \end{array}
    \right]
\end{eqnarray}
so that the solution is
\begin{eqnarray}
  \begin{bmatrix} J(t) \\ \dot J(t)  \end{bmatrix} = E_t
  \begin{bmatrix} J(0) \\ \dot J(0)  \end{bmatrix}
\end{eqnarray}
then we see that the points at $0$ and $t$ are conjugate along the
geodesic if and only if $E_2$ is singular.

Suppose now that we begin at a point $x$ and move along a geodesic.
Let $t_c > 0$ be the first time we pass through a point $y$ conjugate
to $x$, assuming such a point exists.  This point is of particular
interest, because it can be shown (see~\cite{Berger03a}, pp~268-270)
that past the first conjugate point the geodesic is no longer
minimizing.  Thus, the propagator $E_t$ associated with the Jacobi
equation provides a computational machine which lets us determine when
geodesics are no longer minimizing\footnote{Note, however, that while
  the conjugate point condition is sufficient to say a geodesic is no
  longer minimizing, it may not be necessary --- there could already
  be a (globally) shorter path before a conjugate point is
  encountered.}.

To illustrate these ideas we analyze geodesics of the form $e^{-iHt}$,
where $H$ is a sum of one- and two-body terms.  Intuitively, over
short times we expect that the fastest way to simulate a physical
system is with its own evolution.  This intuition is confirmed by the
fact (\cite{Berger03a}, pp~222-226) that over sufficiently short time
periods geodesics are guaranteed to be globally minimal paths.

However, over longer time periods this is no longer the case.  A
simple illustration is the unitary $e^{-iZt}$.  For $t \leq \pi/2$
this can be implemented by applying the Hamiltonian $H = Z$ for a time
$t$; for $\pi > t > \pi/2$ it is more efficient to apply $H= -Z$ for a
time $\pi-t$.  More generally, for any Hamiltonian it is true that
$e^{-iHt} \approx I$ for sufficiently large $t$, and so sufficiently
long geodesics are never minimizing.

The method of conjugate points offers a powerful general way of
studying when geodesics are no longer globally minimal.  In
particular, by integrating the Jacobi equation it is possible to
determine to good accuracy when two points are conjugate using
computational resources that scale polynomially with the dimension of
the underlying manifold, which is $2^{O(n)}$ in this instance.  By
contrast, the volume of the space of paths scales as $2^{O(2^n)}$.  It
is therefore \emph{a priori} quite remarkable that it is possible to
prove a geodesic is not minimizing using $2^{O(n)}$ computational
resources.

In general, finding conjugate points seems to require numerical
solution of the Jacobi equation, perhaps using an explicit solution
such as
Equation~(\ref{eq:explicit_solution_lifted_Jacobi_constant_Ham}),
valid in some special case.  In the bi-invariant case, i.e., when
$q=1$, it is possible to write an analytic solution.  In particular,
Equation~(\ref{eq:explicit_solution_lifted_Jacobi_constant_Ham})
simplifies because $A = 0$, and it is easily verified that
conjugate points occur at times
\begin{eqnarray}
  t_c = \frac{2 m \pi}{\lambda_j-\lambda_k}, \label{eq:conj_points_biinvariant}
\end{eqnarray}
where $m$ is a non-zero integer, and $\lambda_j$ and $\lambda_k$ are distinct
eigenvalues of $H$.

\begin{figure}
\includegraphics[width=85mm]{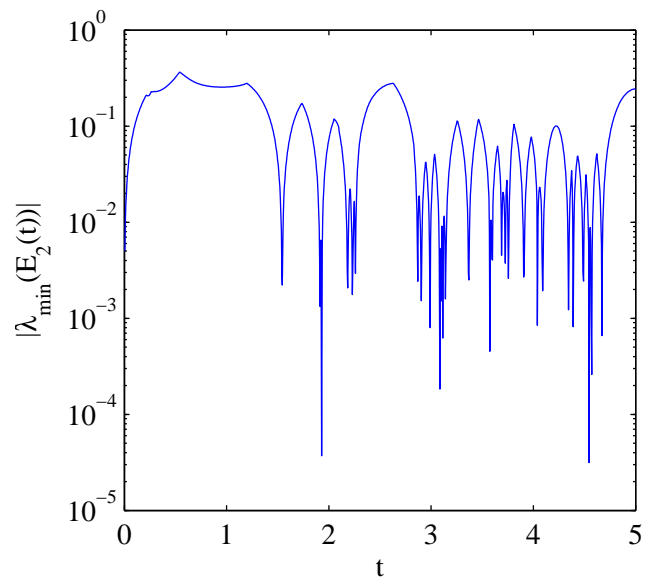}
\caption{Log plot of the absolute value of the minimum eigenvalue of $E_2$ 
  ($\lambda_{\min}(E_2)$) versus time, for the transverse Ising model
  with external field $h=1$.  The penalty parameter is in the regime
  of computational interest, $q=4^n=64$. Sharp dips indicate conjugate
  points.}
  \label{fig:conjugate_example}
\end{figure}

Of course, the case of computational interest is when $q \gg 1$.  As
an example of this case we consider the transverse Ising Hamiltonian
in one dimension,
\begin{eqnarray}
  H = \sum_j Z_jZ_{j+1} + h \sum_j X_j,
\end{eqnarray}
where $h$ is the strength of the applied field.  We numerically
investigated conjugate points for the case of $n = 3$ qubits, with
external field $h = 1$, and penalty $q = 4^n = 64$ chosen to be in the
regime of computational interest.  Figure~\ref{fig:conjugate_example}
is a log plot of the minimum eigenvalue of $E_2$ versus time.  The
occurrence of sharp dips in this plot strongly suggests the presence
of a conjugate point.  We see that in this example the first conjugate
point occurs at $t_c \approx 1.54$, and thus the geodesic is no longer
minimizing past this time.  What is remarkable about this observation
is that we have deduced the non-minimizing property without explicitly
finding a shorter geodesic.  We do it instead through the (relatively)
computationally easy process of studying the conjugate points.  It
would be an interesting challenge to generalize this procedure to
arbitrary $n$, perhaps using the known analytic solution to the
transverse Ising model.

\subsection{Geodesic derivative}
\label{subsec:geodesic_derivative}

Suppose $\gamma(t)$ is a geodesic passing through the point $x$ at
$t=0$ and $y$ at $t=T$, and we vary the metric while holding the
endpoints fixed.  We show in this section that provided $x$ and $y$
aren't conjugate along the geodesic $\gamma$, the geodesic deforms in
a unique way that can be described by an object we call the
\emph{geodesic derivative}.

To define the geodesic derivative, suppose $\gamma(s,t)$ is a family
of geodesics, with $\gamma(s,\cdot)$ being a geodesic for the metric
$g_s$, as in Subsection~\ref{subsec:lifted_Jacobi_equation}.  We
suppose $\gamma(s,t)$ has the constraints $\gamma(0,t) = \gamma(t)$,
and $\gamma(s,0) = x, \gamma(s,T) = y$ for all $s$.  We define the
\emph{geodesic derivative} as $D\gamma \equiv \partial_s \partial_t
\gamma \in T_x M$ at $s= 0$ and $t = 0$.  Note that $\partial_s$ is
defined here, since $\partial_t \gamma$ lives in the same vector
space, $T_x M$, for all values of $s$.  The geodesic derivative thus
represents the way in which the initial tangent $\partial_t \gamma$ is
changing as the parameter $s$ is changing near $s=0$.

\begin{theorem}
  Let $x = \gamma(0)$ and $y = \gamma(T)$ be endpoints on a geodesic
  $\gamma(t)$.  Then a corresponding geodesic derivative $D\gamma$
  exists and is uniquely defined if and only if $x$ and $y$ are not
  conjugate along $\gamma$.
\end{theorem}

Intuitively, at a conjugate point geodesics in a fixed geometry
``split'' into many nearby geodesics (consider, e.g., antipodal points
on a sphere).  Thus, in one direction this theorem is not surprising:
we expect conjugate points to give rise to many different ways to
deform a geodesic as the metric is changed.  The converse, however, is
rather less obvious.

\textbf{Proof:} The existence and uniqueness of such a $D\gamma$ is
equivalent to the existence of a family $\gamma(s,t)$ of geodesics
satisfying the appropriate endpoint conditions, and such that the
geodesic derivative is the \emph{same} for any such family.  Observe
that if such a $D\gamma$ exists, then $D\gamma = \dot J(0)$,
where $J(t)$ is a lifted Jacobi field.  Thus $D\gamma$ exists and is
unique if and only if the lifted Jacobi equation has a unique solution
satisfying $J(0) = J(T) = 0$.  Comparing with the earlier solution to
the lifted Jacobi equation,
Equation~(\ref{eq:solution_lifted_Jacobi_equation}), we see that this
is equivalent to there existing a unique $D\gamma$ satisfying the
constraint:
\begin{eqnarray} \label{eq:geodesic_derivative_equation}
  E_2 D\gamma = P E_t \int_0^t dr \, E_r^{-1} 
  \begin{bmatrix}
    0 \\ C(r) 
  \end{bmatrix},
\end{eqnarray}
where $P$ projects onto the top block in the block representation $K =
\left[{J_1 \atop J_2}\right]$ used in the solution of the lifted
Jacobi equation, Equation~(\ref{eq:solution_lifted_Jacobi_equation}).
Such a unique solution $D\gamma$ exists if and only if $E_2$ is
invertible.  We saw in the last section that this is equivalent to $x$
and $y$ not being conjugate along $\gamma(t)$.  \textbf{QED}

Note that our proof shows more generally that any $D\gamma$ satisfying
Equation~(\ref{eq:geodesic_derivative_equation}) is a valid geodesic
derivative, even when the geodesic derivative is not uniquely defined.
Thus, the analysis of the geodesic derivative is closely tied to
understanding the kernel of $E_2$.

\subsection{Numerically finding geodesics}
\label{subsec:numerical_geodesics}

In this section we explain how the geodesic derivative may be used
to find geodesics reaching a particular desired target unitary, $U$.
The procedure used is to begin by picking a Hamiltonian $H(0)$ which
generates $U$ at some fixed time $T$ along the $q=1$ geodesic.  This
may be done by picking $H(0)$ so that $U = e^{-i H(0) T}$, i.e., by
computing logarithms.  We now vary the parameter $q$ in the family of
metrics $\mathcal{G}_q = \mathcal{P} + q \mathcal{Q}$, causing a
corresponding change $dH_q(0)/dq = D\gamma$ in the initial
Hamiltonian.  Provided the geodesic derivative $D\gamma$ exists and is
unique for a suitable range of values of $q$, we can integrate to
obtain an initial Hamiltonian $H_q(0)$ generating a geodesic
connecting $I$ and $U$, for any desired value of $q$.

To implement this procedure we need to develop a method to compute
$D\gamma$.  We could do this using
Equation~(\ref{eq:geodesic_derivative_equation}), but in the case of
right-invariant metrics on $SU(2^n)$ a more computationally convenient form is
possible, which we now derive.  Recall that $D\gamma$ is defined
to be a value of $\dot J(0)$ such that when $J(0) = 0$, the solution
to the lifted Jacobi equation satisfies $J(T) = 0$, i.e., no variation
occurs at the endpoint as $q$ is varied.  To analyze the values of
$\dot J(0)$ for which this occurs, we examine the solution to the
lifted Jacobi equation for right-invariant metrics more explicitly.
Observe that since $J(0) = 0$ we have
\begin{eqnarray} \label{eq:lifted_Jacobi_intermediate_3}
  J(T) = \int_0^T dt \, \dot J(t) = \int_0^T dt \, U^\dagger(t) K(t) U(t),
\end{eqnarray}
where $K(t) \equiv U(t) \dot J(t) U(t)^\dagger$, as defined in
Subsection~\ref{subsec:specific_lifted_Jacobi_equation}.  The lifted
Jacobi equation, Equation~(\ref{eq:lifted_jacobi_right_invariant}),
has solution
\begin{equation}
K(t) = \mathcal{K}_t(K(0))-\mathcal{K}_t\left(\int_0^t dr \,
  \mathcal{K}_r^{-1}(C(r)) \right) \, , \label{eq:lifted_jacobi_sol_computational}
\end{equation}
where $C$ is the inhomogeneous part of
Equation~\eqref{eq:lifted_jacobi_right_invariant}, and $\mathcal{K}_t$
is the propagator for the homogeneous form of
Equation~(\ref{eq:lifted_jacobi_right_invariant}), i.e., for the
standard (not lifted) Jacobi equation.  The metric derivative along
our family is $\mathcal{G}' = \mathcal{Q}$, and a calculation shows
that
\begin{equation}
C = \mathcal{F}^2(i[\mathcal{P}(H),\mathcal{Q}(H)]).
\end{equation}
Substituting Equation~(\ref{eq:lifted_jacobi_sol_computational}) into
Equation~(\ref{eq:lifted_Jacobi_intermediate_3}), we obtain
\begin{eqnarray}
  J(T) & = & {\cal J}_T(\dot J(0)) \nonumber \\
  & & - \int_0^T dt \, U(t)^\dagger 
  \mathcal{K}_t \left( \int_0^t dr \, \mathcal{K}_r^{-1}(C(r)) \right) U(t),
  \nonumber \\
  & &
\end{eqnarray}
where ${\cal J}_T$ is the propagator that generates the standard (not
lifted) Jacobi field $J_{\rm stand}(T) = {\cal J}_T( \dot J_{\rm
  stand}(0))$, assuming that $J_{\rm stand}(0) = 0$.  Requiring that
$J(T) = 0$ and identifying $dH_q(0)/dq = D\gamma = \dot J(0)$, we
obtain
\begin{eqnarray} \label{eq:geo_deriv_inter}
  & & \frac{dH_q(0)}{dq} \nonumber \\
  & = & {\cal J}_T^{-1} \left[ \int_0^T dt \, U(t)^\dagger
 \mathcal{K}_t \left( \int_0^t dr \, \mathcal{K}_r^{-1}(C(r)) \right) U(t) \right].
\nonumber \\
\end{eqnarray}
This equation can be simplified by observing that
\begin{eqnarray}
&&-\mathcal{K}_t\left(\int_0^t dr \, \mathcal{K}_r^{-1}(C(r)) \right) \label{eq:K_prop} \\ \nonumber 
&=& \left \{\begin{array}{c} i t [\mathcal{Q}(H), \mathcal{P}(H)] \,, 
  \qquad q=1 \,, \\ 
  (\mathcal{K}_t(L(0))-L(t)) / q(q-1) \,, \qquad q>1 \,, \end{array} \right.
\end{eqnarray}
The top equation can be verified by noting that for $q=1$
$\mathcal{K}_t$ is the identity operation for all time and $C$ is
constant.  For the bottom equation it is sufficient to check that both
sides solve Equation~\eqref{eq:lifted_jacobi_right_invariant} with
$q>1$ and initial condition $K(0) = 0$.  Substituting into
Equation~(\ref{eq:geo_deriv_inter}) and using $U^\dagger(t) L(t) U(t)
= L(0)$, we obtain
\begin{eqnarray}
&&\frac{dH_q(0)}{dq} = \label{eq:geo_deriv_computational} \\ \nonumber
&&\left \{\begin{array}{c} \mathcal{J}_T^{-1} \left(\int_0^T dt \, U^\dagger(t) i t [\mathcal{P}(H), \mathcal{Q}(H)] U(t) \right) \,, \qquad q=1 \,, \\ 
(\mathcal{J}_T^{-1}(L(0))T-L(0)) / q(q-1) \,, \qquad q>1 \,, \end{array} 
\right. 
\end{eqnarray}
This is our desired expression for the geodesic derivative.  In
practice, we find it more convenient numerically to work with the
corresponding expression for $dL_q(0)/dq$, which is easily obtained
from this expression using the chain rule.

\textbf{Numerical examples:} We now illustrate the geodesic
deformation procedure for two examples.  The first example is unitary
operations chosen at random, and the second example is the quantum
Fourier transform.

\begin{figure}
\includegraphics[width=85mm]{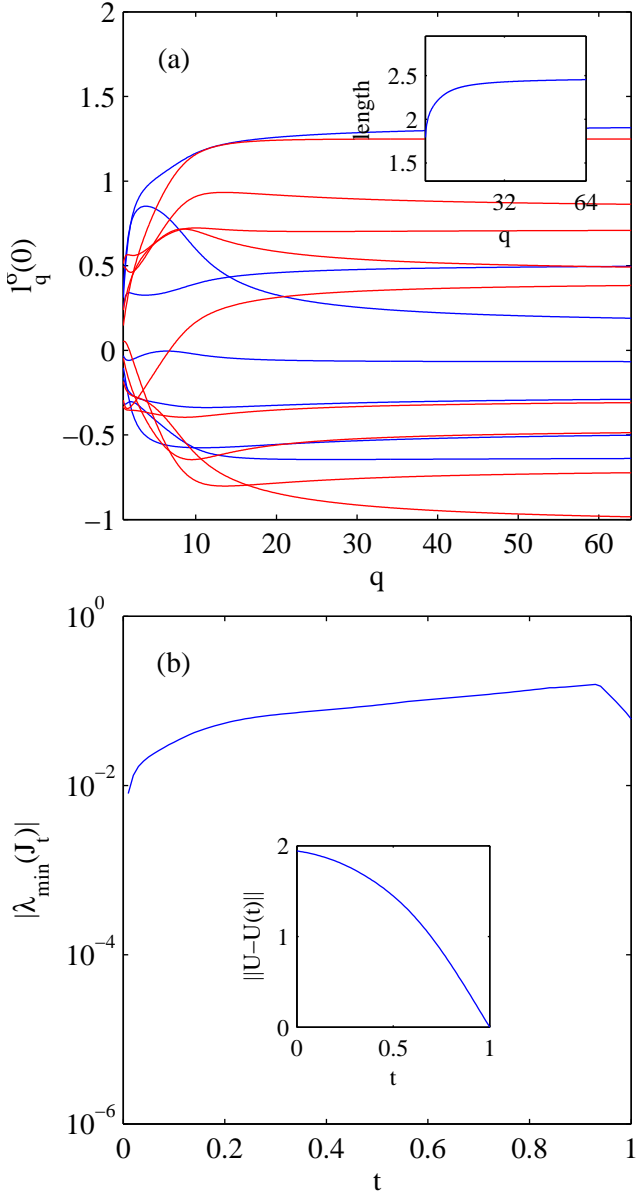}
\caption{Geodesic deformation to a randomly-chosen unitary on $n=3$ qubits.  
  Panel $(a)$ shows how the Pauli components of the initial dual
  Hamiltonian, $l_q^\sigma(0) = \tr(\sigma L_q^\sigma(0))/2^n$, vary
  with the penalty parameter up to $q=4^n=64$.  Of the $4^n=64$ Pauli
  components, only $16$ representatives are shown, for clarity, but all converge in
  the large $q$ limit.  Blue lines are components where $\sigma \in
  \mathcal{P}$, red lines are where $\sigma \in \mathcal{Q}$.  The
  inset shows how the length of the geodesic segment from $I$ to $U$ varies with $q$.  Panel
  $(b)$ shows the minimum eigenvalue of the vectorized form of the
  propagator $\mathcal{J}_t$ as a function of time along the geodesic
  found for $q=64$.  No conjugate points are evident.  The inset shows the
  operator norm of the difference between the target unitary $U$, and
  $U(t)$ along the $q=64$ geodesic, showing that the target is indeed
  reached at the final time $T=1$}
  \label{fig:Urand_deform}
\end{figure}

For the first example, we choose a three-qubit unitary operation, $U$,
according to the Haar measure.  Then we define a unique corresponding
\emph{canonical Hamiltonian}, $H_{\rm canon}$, which satisfies $U =
\exp(-i H_{\rm canon} T)$ and has all eigenvalues in the range
$(-\pi/T,\pi/T]$.  We use this canonical Hamiltonian as our initial
condition, since it has the desirable property that the geodesic
$U(t)=\exp(-i H_{\rm canon} t)$ has no conjugate points before $t=T$,
and thus is a likely candidate for the shortest geodesic through
$U=U(T)$ when $q=1$.  The results obtained when we apply the
deformation procedure are illustrated in
Figure~\ref{fig:Urand_deform}.  Empirically we find that for typical
$U$, if we start with the canonical Hamiltonian and deform to large
values of $q$ we never encounter conjugate points, and so the
deformation is uniquely defined.  This agrees with the general
intuition that conjugate points are rare.  We also empirically observe
(but have not proved) that the value of the dual Hamiltonian $L_q(0)$
converges for large $q$.

Other choices for the starting Hamiltonian are possible by adding
multiples of $2 \pi/T$ to the eigenvalues of the canonical
Hamiltonian.  Thus the set of possible starting Hamiltonians that
reach a desired unitary for $q=1$ at time $t=T$ has the structure of a
(displaced) lattice.  In contrast to the canonical Hamiltonian,
however, our numerical results indicate that conjugate points at $t =
T$ do sometimes occur for some of these other starting Hamiltonians,
and so the deformation procedure is not always well defined.  

Somewhat remarkably, in view of this fact, is that our procedure still
works numerically, even when conjugate points appear at $t=T$.  In
particular, if we take advantage of the fact that numerically the
propagator ${\cal J}_T$ is never exactly singular, then it is still
possible to invert, and we can numerically integrate straight through
the range of values of $q$ where (presumably) a conjugate point
occurs.  Although we do not know how to justify this mathematically,
we find empirically that our algorithm still reaches the desired
target unitary.  It seems likely that what is going on is that our
numerical procedure is picking out one possible way of doing the
deformation.  In principle, of course, it may be that no such
deformation exists, but we have not encountered any circumstance where
this appears to be the case.  An interesting observation is that in
contrast to the canonical case, we find that the initial dual
Hamiltonian, $L_q(0)$, tends not to converge for large $q$, but
continues to grow in norm.

\begin{figure}
\includegraphics[width=85mm]{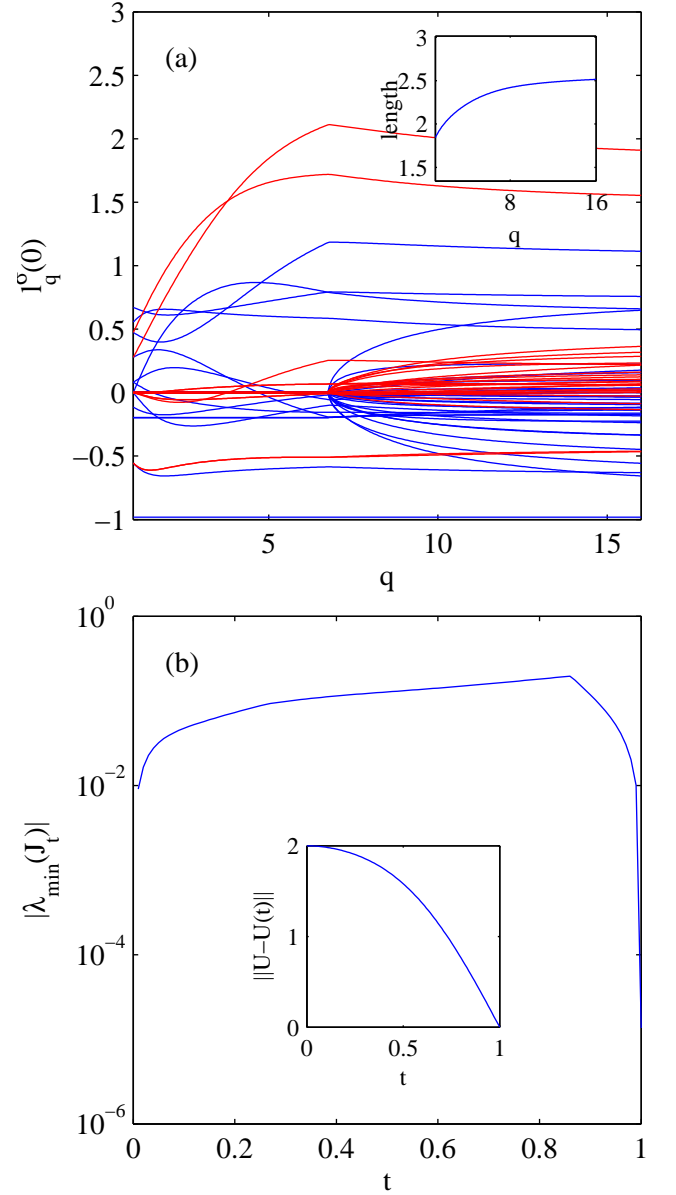}
\caption{Geodesic deformation to the quantum Fourier transform on $n=3$ 
  qubits.  Panel $(a)$ shows how the Pauli components of the initial
  dual Hamiltonian, $l_q^\sigma(0) = \tr(\sigma L_q^\sigma(0))/2^n$,
  vary with the penalty parameter up to $q=16$.  The inset shows how
  the length of the geodesic segment from $I$ to $U$ varies with $q$.
  Blue lines are components where $\sigma \in \mathcal{P}$, red lines
  are where $\sigma \in \mathcal{Q}$.  Panel $(b)$ shows the minimum
  eigenvalue of the propagator $\mathcal{J}_t$ as a function of time
  along the final geodesic found for $q=16$.  The sharp dip at the
  final time $T=1$ indicates a conjugate point.  The inset shows the
  operator norm of the difference between the target unitary $U$, and
  $U(t)$ along the $q=16$ geodesic, showing that the target is indeed
  reached at the final time $T=1$}
  \label{fig:UQFT_deform}
\end{figure}

In our second example, we generate a geodesic reaching the unitary
that implements the quantum Fourier transform on three qubits.  Again
we start with the canonical Hamiltonian for this unitary.  The
deformation is illustrated in Figure~\ref{fig:UQFT_deform}, and only
shows the deformation up to $q=16$ in order to highlight the
interesting behavior around $q=6$.  For values of $q$ in the range $1$
through $\approx 6$ there is a set of Pauli components of $L_q(0)$
that remain zero.  At $q \approx 6$ all of these Pauli components
suddenly become non-zero.  This phenomena coincides with the
propagator $\mathcal{J}_T$ becoming very nearly singular (the
magnitude of its smallest eigenvalue is approximately $10^{-6}$), and
it remains nearly singular up to the final value of $q=16$ as
illustrated in panel $(b)$.  As in the earlier discussion, however, we
find empirically that applying our deformation procedure still appears
to generate a valid (though non-unique) geodesic derivative, and this
is supported by the fact that we do indeed obtain valid geodesics to
the final target unitary, the quantum Fourier transform (see inset of
panel $(b)$).

\textbf{Finding the minimal geodesic:} We have used the geodesic
deformation procedure to obtain upper bounds on the distance $d(I,U)$
for values of $q$ of computational interest.  Of course, there are
many geodesics for any given $U$, not just the geodesics beginning
with the canonical Hamiltonian $H_{\rm canon}$, which is the case we
have focused on.  Can the geodesic deformation procedure be used to
obtain values for the distance $d(I,U)$?

An idea for how to do this is as follows.  Imagine we are trying to
determine whether a geodesic of length $< n^k$ (say) exists for $U$.
It is clear that the length of geodesics monotonically increases with
$q$ under deformation.  At $q=1$ this allows us to restrict our
attention to a finite ($2^{O(n^{k+1})}$) set of possible initial
values for the Hamiltonian, and study how the corresponding geodesics
deform.  A better understanding of the way conjugate points behave
under deformation may enable us to substantially narrow this range of
choices.

A number of caveats to this approach need to be noted.  The first is
that for particular choices of unitary, for example the quantum
Fourier transform above, the deformation procedure produces conjugate
points at the endpoints, and so the deformation procedure is not well
defined.  A possible way around this difficulty is to consider
deforming the metric with two or more parameters, instead of just one.
It seems plausible that it may always be possible to deform the metric
in such a way that new conjugate points never appear along the
geodesics, the intuition being that conjugate points are rather rare.

Secondly, it is possible that as we move to large $q$ new geodesics to
the target unitary appear that cannot be generated as deformations of
a $q=1$ geodesic.  If this is the case then it may be difficult to
deduce anything about the minimal geodesic by deforming.

Despite these caveats, we emphasize that any geodesic between the
identity and the unitary gives an upper bound on the distance of that
unitary from the origin and so is potentially interesting.  For
example, using the methods of~\cite{Nielsen06b}, any geodesic in the
large $q$ limit can be well-approximated by a sequence of one- and
two-qubit quantum gates, and it is plausible that gate sequences
generated in this way may suggest algorithms for computing a family of
unitaries of which the target unitary is a representative (cases of
high symmetry would be natural candidates).  Further study of the
mathematical properties of the geodesic deformation procedure,
including the conjectures mentioned above, is needed to provide more
definitive answers as to its usefulness as a general technique for
finding minimal curves.  It is also desirable to compare to other
techniques (e.g.~\cite{Noakes98a}) which can be used to find geodesics
in spaces of high dimension.

\section{What can geometry teach us about quantum computation?}
\label{sec:discussion}

Motivated by its close connection to quantum gate complexity, in this
paper we have developed a basic understanding of the geometry of the
Riemannian metric defined by Equation~(\ref{eq:standard_metric}).
However, substantial further progress will be required to obtain
either new quantum algorithms or to prove limits on computational
complexity.

In this section we discuss some of the obstacles that need to be
overcome for this to occur.  We begin in
Subsection~\ref{subsec:Razborov-Rudich} with a discussion of the
Razborov-Rudich theorem, a result from classical computational
complexity that illuminates the difficult of proving lower bounds in
classical circuit complexity.  We describe an analogous quantum
result, and an interesting corollary, namely, that if good classical
pseudorandom number generators exist, then the problem of determining
distances on $SU(2^n)$ according to the standard metric is not
(classically) soluble in time polynomial in $2^n$.  The
Razborov-Rudich theorem thus poses a considerable barrier to any
general program for understanding quantum gate complexity, including
our geometric program.  

In Subsection~\ref{subsec:ancilla} we discuss a second obstacle to the
use of geometry, namely that the bounds of
Equation~(\ref{eq:metric_complexity_relationship}) apply only for
circuits which do not make use of ancillary working qubits. While such
circuits are of substantial interest, in general when computing a
desired function $f$ or unitary $U$, it may help to introduce extra
ancillary working qubits.  In this Subsection we explain how ancillas
can be incorporated into the geometric point of view by using a
canonical extension procedure for unitary operations.

\subsection{The Razborov-Rudich theorem and the computational complexity 
  of finding geodesics}
\label{subsec:Razborov-Rudich}

The Razborov-Rudich theorem~\cite{Razborov94a} is a result from
classical computational complexity theory that poses a significant
barrier to any general program for understanding gate complexity,
either classical or quantum.  In this section we briefly survey some
implications the Razborov-Rudich theorem has for the geometric
program.  The discussion is in the nature of an informal outline,
since our intent here is merely to outline the main ideas, rather than
to give the rather extensive formal definitions which a full
discussion would require.  In general, the full formal details are
easy to fill in by experts familiar with the Razborov-Rudich theorem.

In its simplest variant\footnote{See~\cite{Razborov94a} for full
  details.  Our discussion in this paper is for a very simple type of
  natural proof system, in the language of~\cite{Razborov94a}.}, the
Razborov-Rudich theorem shows that, loosely, if good pseudorandom
generators exist, then it is impossible to efficiently distinguish
hard- and easy-to-compute Boolean functions.

This statement can be unpacked in three stages.  First, a ``good''
pseudorandom generator is here taken in the Blum-Micali-Yao sense
(see, e.g., Chapter~9 of~\cite{Arora06a}).  Such generators can be
show to exist if one-way functions exist.  So, for example, if the
factoring or discrete logarithm problems are difficult to solve on a
classical computer, then such generators exist, and the conclusion of
Razborov-Rudich holds.  Second, by hard-to-compute we mean a Boolean
function whose minimal (non-uniform) circuit complexity exceeds some
threshold, e.g., $n^{\ln n}$.  Easy-to-compute means the minimal
circuit complexity is below that threshold.  Note that the threshold
can be varied somewhat, with the result still holding.  Third, by an
efficient procedure to distinguish hard- and easy-to-compute Boolean
functions, we mean an efficient classical Turing machine which takes
as input the truth table for the Boolean function, and determines
whether it is hard- or easy-to-compute.  The criterion for efficiency
is very relaxed: it is that the Turing machine operate in time
polynomial in the size of the truth table, i.e., in time $2^{O(n)}$.

The idea behind the proof of the Razborov-Rudich theorem is easily
stated.  First, observe following Shannon~\cite{Shannon49c} (c.f.
problem 4.4.14 in~\cite{Papadimitriou94a}) that a randomly chosen
Boolean function is with high probability hard to compute.  Second,
using a good pseudorandom number generator it is possible to construct
pseudorandom function generators producing Boolean functions which
appear pseudorandom, but which actually have small circuits.  Any
method for efficiently distinguishing easy- from hard-to-compute
functions would therefore provide an efficient means of distinguishing
random functions from pseudorandom functions, and this contradicts the
definition of a pseudorandom generator.  As a result, such a procedure
cannot exist.

It is straightforward to generalize this reasoning to the quantum
case.  In particular, it can be shown that if good (classical)
pseudorandom generators exist, then there is no efficient classical
algorithm which can be used to distinguish unitary operations that
can be synthesized using small quantum circuits, and those which
require large quantum circuits.  

A similar line of reasoning can be applied to the distance function
$d(I,U)$.  In particular, suppose it were possible to efficiently
distinguish unitaries $U$ for which $d(I,U)$ is large (i.e., exceeds a
threshold like $n^{\ln n})$ from unitaries for which $d(I,U)$ is
small.  Such a procedure could be used to distinguish a unitary chosen
using a pseudorandom generator from one chosen truly at random, and
this could be used to break the pseudorandom generator\footnote{To
  make this description a little more precise, suppose we define $U$
  by $U|x\rangle \equiv (-1)^{f(x)}|x\rangle$, where $f$ is a Boolean
  function that is either generated pseudorandomly, or truly at
  random.}.  It follows that there must be no efficient procedure to
evaluate the distance function $d(I,U)$.

\begin{theorem}
  Suppose classical pseudorandom generators exist. Let $d(I,U)$ be the
  metric on $SU(2^n)$ induced by the standard metric of
  Equation~(\ref{eq:standard_metric}), with $q > 4^n$.  Then there
  is no classical algorithm running in time polynomial in $2^n$ and
  which produces an accurate approximation to $d(I,U)$.
\end{theorem}

This result is particularly remarkable when one considers that when
$q=1$ it \emph{is} possible to evaluate $d(I,U)$ in polynomial time.

These results are, obviously, rather discouraging.  It is worth
emphasizing that analogous results apply to any general approach to
quantum circuit complexity, and are not special to the geometric
approach.  In particular, any approach to the proof of lower bounds
must necessarily contend with the Razborov-Rudich theorem.

Given these results, what is the best approach to finding unitary
operations which can be analyzed using geometric techniques?  We do
not know the answer to this question.  One possibility is to try to
use symmetries to avoid the obstruction posed by Razborov-Rudich.
Symmetries are often used to simplify the analysis of the geodesic
equation, and may, in some cases, make it possible to analyze $d(I,U)$
for those $U$ satisfying the symmetries, without providing a general
efficient procedure for determining $d(I,U)$.  This is currently under
investigation.

\subsection{Extending the geometric picture to take account of ancilla}
\label{subsec:ancilla}

A drawback of the results of~\cite{Nielsen06c,Nielsen06d,Nielsen06b}
is that they apply only to the synthesis of unitary operations without
the assistance of ancillary working qubits.  We now develop a
technique enabling the geometric approach to be applied to many (not
all) unitary operations, even in the case of ancilla.  In particular,
this technique may be applied to unitary operations which compute a
permutation function, $|x\rangle \rightarrow |f(x)\rangle$, or which
are diagonal in the computational basis $|x\rangle \rightarrow
e^{i\theta_x}|x\rangle$.  The technique works by showing that quantum
circuits using ancilla may be put into a standard canonical form which
can then be analyzed geometrically.  

To make the issue at stake more explicit, suppose we wish to
synthesize a unitary operation, $U$, on some number, $n$, of qubits.
To do this synthesis it may help to introduce $m$ additional ancillary
qubits, which start in a standard state.  Without loss of generality
we assume this state is the all $|0\rangle$ state, which we denote
$|0\rangle$.  We then attempt to synthesize a unitary operation $V$
such that for all $n$-qubit states, $|\psi\rangle$,
\begin{eqnarray} \label{eq:ancilla_general_form}
  V|\psi\rangle |0\rangle = (U|\psi\rangle) |A\rangle,
\end{eqnarray}
where $|A\rangle$ is some ancilla state.  Note that by linearity
$|A\rangle$ cannot depend on $|\psi\rangle$.  We call a $V$ satisfying
this relation an \emph{extension} of $U$.  Empirically it is found
that sometimes the gate complexity of synthesizing such an extension
may be strictly less than the gate complexity of synthesizing $U$
without ancilla.

This situation presents a difficulty for the geometric approach, since
it suggests that we need to evaluate the distance $d(I,{\cal U})$,
where ${\cal U}$ is the entire set of extensions of $U$. The set
${\cal U}$ is rather complex, and it seems likely to be far more
difficult to evaluate $d(I,{\cal U})$ than $d(I,U)$.

In this section we show how $d(I,{\cal U})$ can be accurately
estimated using distances $d(I,U')$, for a suitably chosen unitary $U'
= U'(U)$.  This enables us to use distances to obtain bounds on the
gate complexity of unitary operations, with ancilla allowed.  The
constructions we describe do not apply for all unitary operations, but
they do apply for many unitaries of interest, including the unitaries
that arise in the computation of classical functions.

To state our results more formally, let $G_{\infty}(U)$ be the minimal
number of one- and two-qubit gates required to synthesize an extension
of $U$, with an unbounded number of qubits allowed.  We define a
\emph{special extension} of $U$ to be an extension $V$ such that the
final state of the ancilla is the same as the initial state,
$|A\rangle =|0\rangle$, i.e.,
\begin{eqnarray} \label{eq:ancilla_special_form}
  V|\psi\rangle |0\rangle = (U|\psi\rangle) |0\rangle.
\end{eqnarray}
We say a special extension is an \emph{$m$-fold special extension} if
the number of ancilla qubits is $m$. We define $\tilde G_m(U)$ to be
minimal exact gate complexity of an $m$-fold special extension of $U$,
and $\tilde G_{\infty}(U)$ to be the minimal exact gate complexity of
a special extension of $U$ with an unbounded number of ancilla qubits.

We will show how to obtain bounds on $G_\infty(U)$ by showing that for
suitable choice of $m$ there is a single $m+1$-fold special extension
$U_m$ of $U$ such that $G(U_m)$ can be used to bound $\tilde
G_\infty(U)$.  Furthermore, we will also show that for many
interesting unitaries, including all those associated with the
evaluation of classical functions, $G_\infty(U)$ and $\tilde
G_\infty(U)$ behave in essentially the same way.

This allows us to reduce the study of the gate complexity of $U$ with
ancilla to the study of the complexity of a fixed unitary, $U_m$,
without ancilla.  This study can then be done through geometric
methods, or using any other preferred method of analysis.

The bounds relating $\tilde G_\infty(U)$ to $G(U_m)$ and $d(I,U_m)$ go
in one direction.  Bounds in the other direction, analogous to the
first inequality in~(\ref{eq:metric_complexity_relationship}), may be
obtained by replacing $G_\infty(U)$ by an approximate analogue,
$G_\infty(U,\epsilon)$.  

In particular, we define $G_\infty(U,\epsilon)$ to be the minimal
number of one- and two-qubit gates needed to synthesize a unitary
operation $V$ such that $\| {\cal V}-{\cal U}\| \leq \epsilon$, where
${\cal U}$ and ${\cal V}$ are the natural quantum operations on the
$n$-qubit input space induced by $U$ and $V$, and $\| {\cal E} \|
\equiv \max_\rho \mbox{tr} |{\cal E}(\rho)|$, with the maximization
over density matrices $\rho$; note that $\|{\cal V} - {\cal U} \| \leq
\| U - V \|$.

\begin{theorem} {} \label{thm:ancillas}
  There exist positive constants $c_1$ and $c_2$ such that for any $U$
  and $m$ we can construct an $m+1$-fold special extension $U_m$ such
  that:
  \begin{eqnarray}
    \min(m,c_1 G(U_{m})-c_2m) & \leq & \tilde G_\infty(U) \\
    G_\infty(U,\epsilon) & \leq & G(U_m,\epsilon).
  \end{eqnarray}
\end{theorem}

\textbf{Proof:} Let $V$ be any $m$-fold special extension of $U$.
Consider the circuit:
\begin{equation} \label{eq:ancilla_canonical}
\Qcircuit @C=1em @R=.3em @!R {
 \lstick{|x\rangle} & {\hspace{4mm}/} & \multigate{1}{V} & \qw & \multigate{1}{V^\dagger}
 & \qw  & \qw \\
 \lstick{|y\rangle} & {\hspace{4mm}/} & \ghost{V} & \ctrlo{1} & \ghost{V^\dagger}
 & \ctrlo{1} & \qw \\
 \lstick{|z\rangle} & & \ctrlo{-1} & \gate{X} & \ctrlo{-1}
 & \gate{X} & \qw \\
}
\end{equation}
Note that the first wire represents the $n$ qubits on which we desire
to implement $U$, the second wire represents $m$ ancilla qubits, and
the third wire is a single qubit.  Note that operations controlled on
the second wire are only applied if \emph{all} the qubits in the
second wire are set to $|0\rangle$.

We claim that: (1) this circuit defines an $m+1$-fold special
extension of $U$; and (2) the action of this extension is
\emph{independent} of the choice of special extension, $V$, and is
given by the operation $U_m$ defined by the circuit:
\begin{equation} 
U_m = \hspace{5mm}
\begin{array}{l}
\Qcircuit @C=1em @R=.3em @!R {
 & {/}\qw & \gate{U} & \gate{U^\dagger} & \qw \\
 & {/}\qw & \ctrlo{-1} & \ctrlo{-1} & \qw \\
 & \qw & \ctrlo{-1} & \ctrl{-1} & \qw
}
\end{array}
\end{equation}
We call $U_m$ the \emph{$m$'th canonical unitary extension} of $U$.
Note that the fact that $U_m$ is an $m+1$-fold unitary extension of
$U$ follows trivially from the form of this circuit, and so the true
challenge here is to prove that the action of $U_m$ is the same as the
action of the circuit in Equation~(\ref{eq:ancilla_canonical}).  To
verify this it helps to consider separately the cases where $y=0, y
\neq 0$ and $z=0, z=1$.  The three cases (i) $y=0,z=0$, (ii) $y=0, z=
1$ and (iii) $y\neq 0, z=1$ all follow from straightforward circuit
analysis.

The final case, $y\neq 0, z = 0$, requires more care.  After the first
gate is applied, the state is $(V|x\rangle |y\rangle)|0\rangle$.  The
critical claim, proved in the next paragraph, is that the state
$V|x\rangle |y\rangle$ has zero overlap with any state of the form
$|x'\rangle|0\rangle$.  As a result, the second gate has no effect on
the state of the system, and the third gate merely inverts the effect
of the first.  The final gate has no effect (since $y \neq 0$), and
thus the net effect of the circuit is to transform
$|x\rangle|y\rangle|0\rangle$ to $|x\rangle|y\rangle|0\rangle$, which
matches the action of $U_m$.  This completes the proof.

To see that $V|x\rangle |y\rangle$ has zero overlap with any state of
the form $|x'\rangle|0\rangle$, observe that $|x'\rangle|0\rangle = V
(U^\dagger|x'\rangle) |0\rangle$, and thus:
\begin{eqnarray}
  \langle x'|\langle 0| V|x\rangle |y\rangle & = & \langle x'|U \langle 0|
  V^\dagger V |x\rangle|y\rangle \\
  & = & \langle x'|U|x\rangle \langle 0|y\rangle,
\end{eqnarray}
which vanishes since $y \neq 0$.

The remainder of the proof of Theorem~\ref{thm:ancillas} is relatively
straightforward.  The proof that $G_{\infty}(U,\epsilon) \leq
G(U_m,\epsilon)$ follows from the fact that any unitary which
approximates $U_m$ to accuracy $\epsilon$ necessarily approximates $U$
to accuracy $\epsilon$, using standard arguments about operator norms.

To prove the other inequality, note that without loss of generality we
can assume that $m > \tilde G_\infty(U)$ (otherwise the inequality is
trivially true).  In this instance observe that $\tilde G_\infty(U) =
\tilde G_m(U)$, since it cannot help to have more ancilla qubits than
gates in a circuit. Let $V$ be the optimal $m$-fold special extension
of $U$, so $\tilde G_m(U) = G(V)$.  Observe that $U_m$ can be
synthesized using the circuit in
Equation~(\ref{eq:ancilla_canonical}).  It follows that $G(U_m) \leq
c_1 G(V) + c_2 m = c_1 \tilde G_m(U)+c_2m$, where the term liner in
$G(V)$ is due the the controlled-$V$ and -$V^\dagger$, and the term
linear in $m$ is due to the multiply controlled operations.
Rearranging this inequality gives the desired result. \textbf{QED}

The following corollary follows from the theorem and the results of
Section~\ref{sec:background}:
\begin{corollary} \label{cor:ancillas}
  There exist positive constants $c_1$ and $c_2$ such that for any $U$
  and $m$ we can construct an $m+1$-fold special extension, $U_m$ (the
  canonical extension), such that:
  \begin{eqnarray}
    \min(m,c_1 d(I,U_{m})-c_2m) & \leq & \tilde G_\infty(U) \\
    {\rm poly}(G_\infty(U,\epsilon)) & \leq & d(I,U_m).
  \end{eqnarray}
\end{corollary}

In order to apply the theorem and corollary, we need to find scenarios
where $\tilde G_\infty(U)$ and $G_\infty(U)$ behave in essentially the
same way.  We now show that this is the case for Boolean functions, $f
: B_n \rightarrow B$, where $B = \{0,1\}$ is the set of states of a
single bit, and $B_n$ is the set of states of a string of $n$ bits.
An extension to more complex classical functions may be performed
along similar lines.

We define $G_\infty(f)$ to be the minimal number of quantum gates
required to exactly compute $f(x)$.  That is, it is the minimal number
of one- and two-qubit gates required to compute a unitary $V$ such
that:
\begin{eqnarray} \label{eq:circuit_complexity_Boolean_function}
  V|x\rangle |0\rangle = |f(x)\rangle|A_x\rangle,
\end{eqnarray}
where $|A_x\rangle$ is some ``junk'' final state that will be ignored.
We define $G^c_\infty(f)$ to be the minimal classical circuit
complexity required to exactly compute $f$. 

Suppose we define a unitary $U_f$ by $U_f|x\rangle|z\rangle \equiv
|x\rangle|z\oplus f(x)\rangle$, where addition is done modulo two, and
a unitary $V_f$ by $V_f|x\rangle \equiv (-1)^{f(x)}|x\rangle$.  Then
the following theorem shows that the quantum circuit complexity of $f$
is essentially equal to $\tilde G(U_f)$ and $\tilde G(V_f)$.  Thus,
Theorem~\ref{thm:ancillas} and Corollary~\ref{cor:ancillas} may be
applied to obtain insight into the quantum circuit complexity of
Boolean functions.

\begin{theorem} \label{thm:Boolean}
  \begin{eqnarray}
    G^c_\infty(f) \geq G_\infty(f) = \Theta(\tilde G_\infty(U_f)) =
    \Theta(\tilde G_\infty(V_f))
  \end{eqnarray}
\end{theorem}

\textbf{Proof:} The first inequality is obvious.  The first equality
follows by standard techniques of reversible
computation~\cite{Bennett73a,Bennett97b}.  In brief, note that by
definition $G_\infty(f) \leq G_\infty(U_f) \leq \tilde G_\infty(U_f)$.
Conversely, let $V$ be the unitary of minimal gate complexity
satisfying Equation~(\ref{eq:circuit_complexity_Boolean_function}).  Then
by applying $V$ to the first and third register of
$|x\rangle|z\rangle|0\rangle$ we obtain
$|f(x)\rangle|z\rangle|A_x\rangle$.  Adding the value of the first
register to the second and then applying $V^\dagger$ we obtain
$|x\rangle|z\oplus f(x)\rangle|0\rangle$.  It follows that $\tilde
G_\infty(U_f) \leq 2 G_\infty(f)+1$, and thus $G_\infty(f) =
\Theta(\tilde G_\infty(f))$.

The second equality follows by standard techniques of phase
estimation~\cite{Deutsch85a,Cleve98a}; see e.g., Section 5.2
of~\cite{Nielsen00a}. \textbf{QED}

To conclude this section, we give some examples of
Theorem~\ref{thm:ancillas} and its consequences in action.  We will
focus on the behaviour of $\tilde G_\infty(U)$, assuming that we are
working in a situation like that provided by
Theorem~\ref{thm:Boolean}, e.g., with a class of unitaries for which
$\tilde G_\infty(U)$ and $G_\infty(U)$ behave similarly.

A simple example of the theorem is to suppose that $U$ is an $n$-qubit
unitary for which we can prove
\begin{eqnarray} \label{eq:example_assumption}
  d(I,U_{n^2}) \geq \left( \frac{c_2}{c_1}+\delta \right) n^2
\end{eqnarray}
for some $\delta > 0$.  It need not be that $d(I,U_{n^2})$ actually
scales quadratically --- it would be just as good if $d(I,U_{n^2}) =
2^n$, for example.  Substituting $m = n^2$ into the theorem, it
follows that:
\begin{eqnarray}
  \tilde G_\infty(U) \geq \min(1,\delta) n^2 = \Omega(n^2).
\end{eqnarray}
Thus, if $G_\infty(U) \sim \tilde G_\infty(U)$, then we can prove that
the number of gates required to synthesize $U$ scales at least as
$\Omega(n^2)$.

This conclusion perhaps appears somewhat surprising.  After all,
$U_{n^2}$ involves $n^2$ qubits, and so surely we would expect
$d(I,U_{n^2})$ to scale as in Equation~(\ref{eq:example_assumption}),
no matter what $U$ is.  The resolution is that it is only the
\emph{excess} beyond $(c_2/c_1) n^2$ that contributes to the bound on
the gate complexity.  Fortunately, there are many situations where
such an excess is likely to occur.  To see this, recall from
Equation~(\ref{eq:metric_complexity_relationship}) that
\begin{eqnarray} \label{eq:refined_estimate}
  \frac{b_0 G(U,\epsilon)^{b_1} \epsilon^{b_2}}{n^{b_3}} \leq d(I,U).
\end{eqnarray}
In the papers~\cite{Nielsen06c,Nielsen06d} the constants found were
$b_1 = 1/3, b_2 = 2/3$ and $b_3 = 2$; a value for $b_0$ was not
calculated explicitly.  It is now straightforward to prove that:

\begin{proposition}
  \begin{eqnarray}
   \frac{b_0 G_\infty(U,\epsilon)^{b_1} \epsilon^{b_2}}{(n+m)^{b_3}}
   \leq d(I,U_m).
  \end{eqnarray}
\end{proposition}

\textbf{Proof:} Simply observe that $G_\infty(U,\epsilon) \leq
G(U_m,\epsilon)$, and then apply~(\ref{eq:refined_estimate}).
\textbf{QED}

This proposition shows that rapid scaling in $G_\infty(U,\epsilon)$
implies rapid scaling in $d(I,U_m)$.  As a result, if $U$ is difficult
to approximate, then the geometric properties imply that $U$ is
difficult to compute exactly, \emph{even when ancilla are allowed}.
Needless to say, if $U$ is difficult to approximate, then it is
difficult to compute exactly.  The significance of the Proposition is
that it provides circumstances under which we can guarantee something
about the behaviour of the geometry.

As an example, suppose we define $a \equiv (1+b_3)/b_1$, and that
$G_\infty(U,1/10) = \Omega(n^{a+\delta})$ for some $\delta > 0$.
Suppose we choose $\gamma > 0$ such that $\gamma < \delta / a$.  Then
applying the proposition we see with a little algebra that
$d(I,U_{n^{1+\gamma}}) = \Omega(n^{1+\beta})$ for some $\beta >
\gamma$.  In such a situation, it follows from
Theorem~\ref{thm:ancillas} that the geometric properties imply
superlinear lower bounds on the exact gate complexity $\tilde
G_\infty(U)$.

In a similar vein, if we have $G_\infty(U,1/10) = \Omega(2^{c_1 n})$,
and choose a positive value for $c_2$ such that $c_2 < c_1 / a$ then
we see from the proposition that $d(I,U_{2^{c_2n}}) =
\Omega(2^{c_3n})$ for some $c_3 > c_2$, and thus by
Theorem~\ref{thm:ancillas} the geometric properties imply exponential
lower bounds on the exact gate complexity $\tilde G_\infty(U)$.

So, for example, if, as suspected by many people, it turns out that
\textbf{NP}-hard problems require exponential size quantum circuits to
approximate, then it will immediately follow that there are constants
$0 < c_2 < c_3$ such that $d(I,U_{2^{c_2 n}}) = \Omega(2^{nc_3})$, and
thus, by Theorem~\ref{thm:ancillas} the geometric properties imply
exponential lower bounds on the exact gate complexity.

\section{Conclusion}
\label{sec:conclusion}

We have explored the basic geometry of quantum computation, including
the Levi-Civita connection, the geodesic equation and many solutions
and invariants of the equation, as well as all the basic curvature
quantities.  We have also developed a geodesic deformation procedure
which in many cases of interest allows us to find geodesics connecting
the identity $I$ to some desired unitary $U$.  This gives a more or
less complete picture of the basic geometry of quantum computation,
and should provide a foundation for a more detailed understanding.

\acknowledgments

Thanks to Scott Aaronson, Ben Andrews, Ike Chuang, Andrew Doherty,
Mile Gu, and Lyle Noakes for assistance and encouragement.  Steve
Flammia and Bryan Eastin's Qcircuit package was used in the
preparation of this paper.

\appendix

\section{Curvature}
\label{sec:curvature}

In this appendix we derive explicit expressions for the various
quantities describing curvature.  This includes the curvature tensor
(Subsection~\ref{subsec:curvature_tensor}), the sectional curvature
(Subsection~\ref{subsec:sectional_curvature}), the Ricci tensor
(Subsection~\ref{subsec:Ricci_tensor}), and the scalar curvature
(Subsection~\ref{subsec:scalar_curvature}).  Note that many of these
quantities are presented for a general right-invariant metric
in~\cite{Milnor76a}.  However, it is helpful to have explicit forms of
these curvature quantities for the standard metric,
Equation~(\ref{eq:standard_metric}), and so we present detailed
calculations.

\subsection{Curvature tensor}
\label{subsec:curvature_tensor}

In this section we compute the Riemann curvature tensor, $R$, which is
a ${0 \choose 4}$ tensor field defined by $R(W,X,Y,Z) \equiv \langle
\nabla_W \nabla_X Y-\nabla_X \nabla_W Y - \nabla_{i[W,X]} Y,Z
\rangle$. We work with respect to a basis of right-invariant frame
fields $\rho, \sigma, \tau, \mu$, corresponding to generalized Pauli
matrices, and compute the corresponding components of the curvature
tensor,
\begin{eqnarray}
  R_{\rho \sigma \tau \mu} & \equiv & 
  \langle \nabla_\rho \nabla_\sigma \tau - \nabla_\sigma \nabla_\rho \tau
  - \nabla_{i [\rho,\sigma]} \tau, \mu \rangle \\
  & = & \langle \nabla_\rho \tau, \nabla_\sigma \mu \rangle -
  \langle \nabla_\sigma \tau, \nabla_\rho \mu \rangle
  - \langle \nabla_{i[\rho,\sigma]} \tau, \mu \rangle, \nonumber \\
  & & \label{eq:curvature_intermediate}
\end{eqnarray}
where in the second line we used the fact that $\langle X, \nabla_{Y}
Z \rangle = -\langle \nabla_{Y} X, Z \rangle$ for any triple of
right-invariant vector fields, $X, Y$ and $Z$.  Using the formula of
Equation~(\ref{eq:connection_right_invariant}), we see that
\begin{eqnarray} \label{eq:connection_Pauli}
  \nabla_\sigma \tau = i c_{\sigma,\tau} [\sigma,\tau],
\end{eqnarray}
where
\begin{eqnarray} \label{eq:def_c}
  c_{\sigma,\tau} \equiv \frac{1}{2} \left( 1 +\frac{q_{\tau}-q_{\sigma}}
    {q_{[\sigma,\tau]}}\right).
\end{eqnarray}
Note that we use $q_\sigma$ to refer to the value of the penalty for
the generalized Pauli matrix $\sigma$, i.e., $q_\sigma = 1$ if
$\sigma$ has weight zero, one or two, and otherwise $q_\sigma = q$.
The notation $q_{[\sigma,\tau]}$ is the value of the penalty for the
generalized Pauli matrix proportional to $[\sigma,\tau]$; in the
trivial case when $\sigma$ and $\tau$ commute, we arbitrarily assign
$q_{[\sigma,\tau]} = 1$.  We use a similar convention for expressions
like $c_{[\rho,\sigma],\tau}$.

Substituting Equation~(\ref{eq:connection_Pauli}) into
Equation~(\ref{eq:curvature_intermediate}), we obtain
\begin{eqnarray} \label{eq:curvature_tensor}
  R_{\rho \sigma \tau \mu} & = & c_{\rho,\tau} c_{\sigma,\mu}
  \langle i [\rho,\tau], i [\sigma,\mu] \rangle \nonumber \\
  & & - c_{\sigma,\tau}c_{\rho,\mu} \langle i[\sigma,\tau], i[\rho,\mu]\rangle
  \nonumber \\
 & &
  -c_{[\rho,\sigma],\tau} \langle i [i[\rho,\sigma],\tau],\mu \rangle.
\end{eqnarray}
We use this as our basic expression for the curvature tensor, and
derive other curvature quantities starting from this point.  In doing
so we will often find it helpful to use the observation that $\langle
\sigma,\tau \rangle = q_{\sigma} \delta_{\sigma \tau}$.  Note that
this expression has several symmetries in addition to those satisfied
in general by the curvature tensor.  In particular, it is easy to
verify that to have $R_{\rho \sigma \tau \mu} \neq 0$ we must have
$\rho \sigma \tau \mu$ proportional to the identity, and it must be
possible to partition the indices into two pairs, e.g.,
$(\rho,\sigma)$ and $(\tau,\mu)$, such that: (1) the pairs commute,
i.e., $[\rho, \sigma ] = [\tau,\mu] = 0$; and (2) all other pairs
anticommute, i.e., $[\rho,\tau]_+ = [\rho,\mu]_+ = \ldots = 0$.  Even
when these conditions hold, individual terms in the
expression~(\ref{eq:curvature_tensor}) may still vanish, e.g., the
first term vanishes if $\rho$ and $\tau$ commute, or if $\sigma$ and
$\mu$ commute.

\subsection{Sectional curvature}
\label{subsec:sectional_curvature}

The sectional curvature in the tangent plane spanned by orthonormal
right-invariant vector fields $X$ and $Y$ is defined by
\begin{eqnarray}
  K(X,Y) \equiv R(X,Y,Y,X).
\end{eqnarray}
Define a bilinear operation ${\cal B}(X,Y) \equiv {\cal F}(i[{\cal
  G}(X),Y])$.  Observe that we have the identities $\langle X, i[Y,Z]
\rangle = \langle {\cal B}(X,Y), Z \rangle$ and $\nabla_X Y =
\frac{1}{2} ( i [X,Y]-{\cal B}(X,Y)-{\cal B}(Y,X))$.  Using these
facts and the cyclic property of trace a calculation shows that
\begin{eqnarray} \label{eq:sectional_curvature}
  K(X,Y) & = & - \frac{3}{4} \langle i[X,Y],i [X,Y] \rangle \nonumber \\
 & &
    + \frac{1}{4} \langle {\cal B}(X,Y)+{\cal B}(Y,X),  {\cal B}(X,Y)+{\cal B}(Y,X) \rangle \nonumber \\
    & & + \frac{1}{2} \langle i [X,Y], {\cal B}(X,Y)-{\cal B}(Y,X) \rangle.
\end{eqnarray}
For values of $q$ of computational interest (indeed, for any $q >
4/3$) it is easily verified from this formula that the sectional
curvature can be both positive and negative.

\subsection{Ricci tensor}
\label{subsec:Ricci_tensor}

The Ricci tensor, $Rc_{\sigma \tau}$, is defined as the contraction of
the raised form of the curvature tensor, $R_{\rho \sigma
  \tau}^{\phantom{\rho \sigma \tau} \mu}$, on the first and last
indices,
\begin{eqnarray}
Rc_{\sigma \tau} & = & R_{\rho \sigma \tau}^{\phantom{\rho \sigma \tau} \rho}
\\
& = & g^{\rho \mu} R_{\rho \sigma \tau \mu}.
\end{eqnarray}
Observing from Equation~(\ref{eq:inverse_metric_components}) that
$g^{\rho \mu}$ is nonzero only when $\rho = \mu$, and that $R_{\rho
  \sigma \tau \mu}$ vanishes unless $\rho \sigma \tau \mu \propto I$,
we see that $Rc_{\sigma \tau}$ must be diagonal, i.e., the components
vanish unless $\sigma = \tau$.

To compute the diagonal entries in the Ricci tensor, we observe from
Equation~(\ref{eq:inverse_metric_components}) that the diagonal
entries of the metric $g^{\rho \tau}$ are equal to $1/q_{\rho}$.  We
thus obtain (no implied sum on repeated indices)
\begin{eqnarray}
  Rc_{\sigma \sigma} = \sum_\rho \frac{R_{\rho \sigma \sigma \rho}}{q_{\rho}}.
\end{eqnarray}
Using the expression of Equation~(\ref{eq:curvature_tensor}), the
definition Equation~(\ref{eq:def_c}), and the observations $\langle
i[\rho,\sigma], i [\rho,\sigma]\rangle = 4 q_{[\rho,\sigma]}$ and
$\langle i [i[\rho,\sigma],\sigma],\rho\rangle = -4 q_\rho$, we
obtain after some algebra
\begin{eqnarray}
Rc_{\sigma \sigma} =
\sum_\rho' \left( 2+\frac{q_\rho^2+q_\sigma^2-2 q_\rho q_\sigma -3 q_{[\rho,\sigma]}^2 
  + 2 q_{[\rho,\sigma]} q_\sigma}{q_\rho q_{[\rho,\sigma]}} \right), \nonumber \\
\end{eqnarray}
where the prime indicates that the sum is only over $\rho$ which
anticommute with $\sigma$.  This sum may be further simplified by
observing that up to proportionality factors $\rho$ and
$[\rho,\sigma]$ range over the same set of matrices, i.e., generalized
Pauli matrices which anticommute with $\sigma$.  Using this fact a
change of variables may be used to show that the sum of the $-2 q_\rho
q_\sigma/ q_\rho q_{[\rho,\sigma]}$ and $2 q_{[\rho,\sigma]} q_\sigma/
q_\rho q_{[\rho,\sigma]}$ terms cancel.  For similar reasons, the
$q_\rho^2 / q_\rho q_{[\rho,\sigma]}$ and $-3 q_{[\rho,\sigma]}^2/
q_\rho q_{[\rho,\sigma]}$ terms partially cancel.  Finally, provided
$\sigma \neq I$, a simple counting argument shows that the number of
$\rho$ which anticommute with $\sigma$ is $4^n/2$.  Combining all
these observations, we obtain
\begin{eqnarray} \label{eq:Ricci_explicit}
Rc_{\sigma \sigma} =
4^n + \sum_\rho' \frac{q_\sigma^2-2 q_\rho^2}{q_\rho q_{[\rho,\sigma]}}.
\end{eqnarray}

To evaluate this more explicitly, let us define $N_\sigma({\cal
  P},{\cal P})$ to be the number of generalized Pauli matrices $\rho$
such that: (1) $\rho$ anticommutes with $\sigma$; (2) $\rho$ is in
${\cal P}$, i.e., has only one- or two-body terms; and (3)
$[\rho,\sigma]$ is also in ${\cal P}$.  We make analogous definitions
for $N_\sigma({\cal P},{\cal Q}), N_\sigma({\cal Q},{\cal P})$ and
$N_\sigma({\cal Q},{\cal Q})$.  Expressed in these terms we have:
\begin{eqnarray}
Rc_{\sigma \sigma} & = &
4^n + q_\sigma^2 \left(N_\sigma({\cal P},{\cal P})+
  \frac{1}{q}N_\sigma({\cal P},{\cal Q}) \right. \nonumber \\
& & \left. + \frac{1}{q} N_\sigma({\cal Q},{\cal P})
  + \frac{1}{q^2}N_\sigma({\cal Q},{\cal Q}) \right) \nonumber \\
& & -2\left( N_\sigma({\cal P},{\cal P})
  +\frac{1}{q} N_\sigma({\cal P},{\cal Q})  \right. \nonumber \\
  & & \left. \vphantom{\frac{1}{q}} +q N_\sigma({\cal Q},{\cal P})
  + N_\sigma({\cal P},{\cal P}) \right).
\end{eqnarray}
Elementary counting allows us to evaluate the factors
$N_\sigma(\cdot,\cdot)$.  It is most convenient to consider separately
the cases where the weight $w$ of $\sigma$ is $1, 2, 3$ and $4$ or
more.  The corresponding values for $Rc_{\sigma \sigma}$ are
\begin{eqnarray}
  & & \wt(\sigma) = 1: Rc_{\sigma \sigma} = 
  2(3n-2)+\left( \frac{4^n}{2}-2(3n-2) \right) \frac{1}{q^2} \nonumber \\
  & & \label{eq:Rc_1} \\
  & & \wt(\sigma) = 2: Rc_{\sigma \sigma} = 
 -24(n-2)q +8(6n-11) \nonumber \\
 & & \phantom{\wt(\sigma) = 2: Rc_{\sigma \sigma} = }
 + \left(\frac{4^n}{2}-8(3n-5)
    \right)\frac{1}{q^2} \label{eq:Rc_2} \\
  & & \wt(\sigma) = 3: Rc_{\sigma \sigma} = 12 q^2
  + \frac{4^n}{2}+36 (n-3) \nonumber \\
  & & \phantom{\wt(\sigma) = 3: Rc_{\sigma \sigma} =}
  -12(3n-8)\frac{1}{q} \label{eq:Rc_3} \\
  & & \wt(\sigma) = w \geq 4: Rc_{\sigma \sigma} = 
  \frac{4^n}{2}+4w(3n-2w) \nonumber \\
  & & \phantom{\wt(\sigma) = w \geq 4: Rc_{\sigma \sigma} = } 
  -4w(3n-2w)\frac{1}{q}. \label{eq:Rc_4}
\end{eqnarray}

\textbf{The Ricci flow:} In Section~\ref{sec:geodesic_deformation} we
study the way geodesics deform when the metric is smoothly changed.  A
well-known method for changing the metric is the \emph{Ricci flow}
introduced by Hamilton~\cite{Hamilton82a} and recently used by
Perelman~\cite{Perelman02a,Perelman03a,Perelman03b} in the resolution
of the Poincare conjecture.  The normalized Ricci flow is an equation
for the metric tensor defined in components by $\partial g_{\sigma
  \tau}/\partial s = -2R_{\sigma \tau}+2 R g_{\sigma \tau}/(4^n-1)$.
This equation defines a smooth family $g_s$ of metrics on the
manifold.  Although we do not seriously study the Ricci flow in this
paper, we now briefly digress to note some interesting properties of
the behaviour of the standard metric under the Ricci flow.  Our
numerical investigations suggest that the normalized Ricci flow takes
the standard metric to the bi-invariant metric with $q=1$, up to an
overall scaling factor.  This is interesting, and deserves further
study, for the geodesics of the bi-invariant metric are well
understood.

To understand the normalized Ricci flow, observe from
Equation~(\ref{eq:Ricci_explicit}) that if $q_\sigma$ depends only on
the weight of $\sigma$, then the resulting diagonal entries
$Rc_{\sigma \sigma}$ in the Ricci tensor depend only on the weight of
$\sigma$.  As a result, under the Ricci flow we can assume that the
metric tensor is always diagonal with entries that depend only on the
weight.

To obtain an explicit expression for the metric under the Ricci flow,
we define $N_\sigma(v,w)$ to be the number of $\rho$ with weight $v$
such that the commutator $[\sigma,\rho]$ is nonvanishing with weight
$w$.  Equation~(\ref{eq:Ricci_explicit}) can be rewritten
\begin{eqnarray}
  Rc_{\sigma \sigma} =  4^n + \sum_{vw} N_\sigma(v,w)
    \frac{q_\sigma^2-2q_v^2}{q_v q_w},
\end{eqnarray}
where $q_v$ is the penalty for Pauli matrices of weight $v$.  To find
a simple formula for $N_\sigma(v,w)$, we observe that $N_\sigma(v,w) =
3^v {n \choose v} p(w|\sigma,v)$, where $p(w|\sigma,v)$ is the
conditional probability that a random Pauli of weight $v$ will commute
with $\sigma$ to give a Pauli of weight $w$.  This probability can be
computed by conditioning on the size of the overlap between $\sigma$
and the Pauli of weight $v$.  The probability is zero unless
$\wt(\sigma)+v-w$ is an odd and positive number.  If that is the case
then
\begin{eqnarray}
  N_\sigma(v,w) & = & \frac{3^v}{2^{\wt(\sigma)+v-w}}
  \sum_{k} \left( \frac{4}{3} \right)^k
  {\wt(\sigma) \choose k} {n-\wt(\sigma) \choose v-k} \nonumber \\
  & & \hspace{2cm} 
  \times {k \choose \wt(\sigma)+v-w-k},
\end{eqnarray}
where $k$ runs over the possible sizes of the overlap region.  A
similar calculation can be done to obtain an expression for the scalar
curvature.  These provide elegant expressions for the Ricci tensor and
scalar curvature in the cases when the metric is diagonal with entries
depending only on the weight, and are useful in numerically simulating
the normalized Ricci flow.

\subsection{Scalar curvature}
\label{subsec:scalar_curvature}

Returning to the study of the standard metric, the scalar curvature is
obtained from the Ricci tensor via the contraction $R =
Rc_\sigma^{\phantom{\sigma} \sigma} = \sum_\sigma Rc_{\sigma \sigma} /
q_\sigma$.  Using Equations~(\ref{eq:Rc_1})-(\ref{eq:Rc_4}) we obtain
\begin{eqnarray}
  R & = & -54 n (n-1) (n-2) q +6n(36n^2-99n+64)  \nonumber \\
  & & +\left[
  \left(4^n-1+\frac{3n(3n-1)}{2}\right)\frac{4^n}{2} \right. \nonumber \\
 & & \left. \phantom{\frac{4^n}{2}} -6n(45n^2-117n+74)\right] \frac{1}{q}
\nonumber \\
  & & -\left[ 3n(3n-1)4^{n-1} -6n(3n-4)(6n-7)\right] \frac{1}{q^2} \nonumber \\
\end{eqnarray}
For large $n$ and fixed $q$ the dominant terms in the scalar curvature
are therefore
\begin{eqnarray} \label{eq:scalar_asymptotic}
  R \sim -54n^3 q + 216 n^3 + \frac{16^n}{2} \frac{1}{q} 
  -9 n^2 4^{n-1} \frac{1}{q^2}.
\end{eqnarray}
We see that provided $q \sim 4^n$, the scalar curvature is necessarily
negative.  Remarkably, the proof in~\cite{Nielsen06c} that
Equation~(\ref{eq:metric_complexity_relationship_2}) holds also
requires $q \sim 4^n$ (or larger), and thus entails negative scalar
curvature.  Whether a relationship like
Equation~(\ref{eq:metric_complexity_relationship_2}) can be proved for
smaller values of $q$ (and thus for positive scalar curvature) remains
an open question.

The scalar curvature can be shown to be the average of the sectional
curvature,
\begin{eqnarray} \label{eq:average_curvature}
  R = (4^n-1) \int d\mu(X,Y) K(X,Y),
\end{eqnarray}
where $\mu(X,Y)$ is the normalized measure induced by our metric on
the space of orthonormal $X$ and $Y$.  Note that the constant of
proportionality out the front is $4^n-1$ if we are working on
$U(2^n)$, and is $4^n-2$ if we are working on $SU(2^n)$.  This
suggests (and Equation~(\ref{eq:sectional_curvature}) can be used to
verify) that typical values of the sectional curvature are
negative.  It is well known that on manifolds with everywhere negative
curvature, the dynamical system defined by the geodesic flow is
ergodic and mixing; see Sections~10.5 and~10.6 of~\cite{Berger03a} for
an overview and references.  This suggests the conjecture that such
ergodic and mixing behaviour may be seen at least on parts of our
manifold.  If true, this may have interesting implications for quantum
computation.

\end{document}